\documentclass[reprint,superscriptaddress,amsmath,amssymb,aps,prl]{revtex4-2}
\usepackage{graphicx}
\usepackage{dcolumn}
\usepackage{bm}
\usepackage{color}
\usepackage{units}
\usepackage{wasysym}
\usepackage{MnSymbol}
\usepackage{comment}
\usepackage{times}
\usepackage{upgreek}
\usepackage [english]{babel}
\newcommand{\fakesection}[1]{
  \par\refstepcounter{section}
  \sectionmark{#1}
  \addcontentsline{toc}{section}{\protect\numberline{\thesection}#1}
}
\usepackage [autostyle, english = american]{csquotes}
\MakeOuterQuote{"}
\usepackage[mathlines]{lineno}
\usepackage[unicode=true,pdfusetitle,bookmarks=false,colorlinks=true,citecolor=blue,urlcolor=blue,linkcolor=blue]{hyperref}
\begin{document}

\title{Phase Separation, Capillarity, and Odd Surface Flows in Chiral Active Matter}

\author{Luke Langford}
\affiliation{Department of Materials Science and Engineering, University of California, Berkeley, California 94720, USA}

\author{Ahmad K. Omar}
\email{aomar@berkeley.edu}
\affiliation{Department of Materials Science and Engineering, University of California, Berkeley, California 94720, USA}
\affiliation{Materials Sciences Division, Lawrence Berkeley National Laboratory, Berkeley, California 94720, USA}

\begin{abstract}
Active phase separations evade canonical thermodynamic descriptions and have thus challenged our understanding of coexistence and interfacial phenomena. 
Considerable progress has been made towards a non-equilibrium theoretical description of these traditionally thermodynamic concepts.
Spatial parity symmetry is conspicuously assumed in much of this progress, despite the ubiquity of chirality in experimentally realized systems. 
In this Letter, we derive a theory for the phase coexistence and interfacial fluctuations of a system which microscopically violates spatial parity.
We find suppression of the phase separation as chirality is increased as well as the development of steady-state currents tangential to the interface dividing the phases.
These odd flows are irrelevant to stationary interfacial properties, with stability, capillary fluctuations, and surface area minimization determined entirely by the capillary surface tension.
Using large-scale Brownian dynamics simulations, we find excellent agreement with our theoretical scaling predictions.
\end{abstract}

\maketitle

\textit{Introduction.--} 
Active phase separation, a phenomenon displayed by a broad array of systems ranging from the internals of a living cell~\cite{Brangwynne2009GermlineDissolution/Condensation} to self-catalytic colloidal surfers~\cite{Palacci2013LivingSurfers} and motile bacteria~\cite{Liu2019Self-DrivenFormation}, has been a subject of considerable recent interest in statistical physics, biology, and soft matter.
Both the phase behavior and interface dividing the coexisting phases has been subject of many intriguing findings ranging from odd surface flows~\cite{Soni2019TheFluid} to propagating waves~\cite{Adkins2022}.
In particular, the interface dividing motility induced phase separations (MIPS)~\cite{Cates2015} has been associated with a \textit{negative} mechanical surface tension~\cite{Bialke2015}, which has sparked much controversy~\cite{Hermann2019b,Omar2020,Li2023}.
Considerable recent progress has been made towards the derivation of physically relevant surface tensions in phase separated active systems~\cite{Lee2017InterfaceSeparations,Patch2018,Fausti2021CapillarySeparation,Langford2023}.
Despite this progress, much of the theoretical insights have tacitly \textit{assumed dynamics respecting spatial parity} -- a symmetry that many active systems violate. 
Whether arising from externally applied magnetic fields~\cite{Soni2019TheFluid}, core biological behavior~\cite{Tan2022OddCrystals}, or a helical nanomotor geometry~\cite{Mandal2018MagneticPropulsion}, these systems are inherently \textit{chiral}.
While the theory surrounding \textit{odd transport} phenomena that generically arises from chiral active systems is increasingly well-established~\cite{Kummel2013CircularParticles,Banerjee2017OddFluids,Hargus2020TimeResults,Han2021FluctuatingFluids,Hargus2021OddMotion,Hargus2024ThePhenomena}, the \textit{phase behavior} and \textit{interfacial} properties of such systems has remained relatively unexplored. 

In this Letter, we systematically derive the interfacial dynamics of a chiral active interface beginning from microscopic dynamics that violates \textit{both} time reversal symmetry and spatial parity.
En route to doing so, we develop a nonequilibrium mechanical description~\cite{Omar2023b} of chiral MIPS, finding that chirality raises the critical motility for the onset of phase separation.
Building on a number of frameworks~\cite{Bray1994TheoryKinetics,Bray2002InterfaceShear,Fausti2021CapillarySeparation,Langford2023}, we then derive an approximate fluctuating hydrodynamic description of interacting chiral active particles and ultimately the dynamics of a phase-separated chiral active interface.
These dynamics reveal the microscopic definition for the capillary tension $\upgamma_{\rm cw}$, which governs the stability of the interface, as well as the odd flow coefficient $\nu_{\rm odd}$ which generates travelling waves at the surface. 
Surprisingly, we find that the stationary statistics of the interface are controlled \textit{exclusively} by $\upgamma_{\rm cw}$ while $\nu_{\rm odd}$ only affects dynamical correlations.
Brownian dynamics simulations of chiral active matter allow us to confirm these theoretical predictions and the precise activity and chirality dependencies of the observed phenomena.

\textit{Model System.--}
We consider a \textit{two-dimensional} (2D) system of $N$ interacting chiral active Brownian particles (cABPs)~\cite{Liao2018ClusteringMonolayer,Ma2022DynamicalParticles,Bickmann2022AnalyticalSwimmers,Batton2024MicroscopicEnvironments} in which the time variation of the position $\mathbf{r}_i$ and (2D) orientation $\mathbf{q}_i$ of the $i$th particle follow overdamped Langevin equations:
\begin{subequations}
  \label{eq:eom}
    \begin{align}
        \Dot{\mathbf{r}}_i &= U_o \mathbf{q}_i + \frac{1}{\zeta}\sum_{j\neq i}^N\mathbf{F}_{ij}\label{eq:rdot}, \\ \Dot{\mathbf{q}}_i &= \boldsymbol{\omega}_o\times\mathbf{q}_i +  \bm{\Omega}_i\times\mathbf{q}_i\label{eq:qdot},
    \end{align} 
\end{subequations}
where $\mathbf{F}_{ij}$ is the (pairwise) interparticle force; $U_o,\boldsymbol{\omega}_o$ are the intrinsic active speed and angular velocity, respectively; $\zeta$ is the translational drag coefficient; and $\bm{\Omega}_i$ is a stochastic angular velocity with zero mean and variance $\langle \bm{\Omega}_i (t) \bm{\Omega}_j(t')\rangle = 2D_R\delta_{ij}\delta(t-t')\mathbf{I}$. 
Here, $D_R$ is the rotational diffusivity, which may be athermal in origin, and is inversely related to the orientational relaxation time $\tau_R \equiv D_R^{-1}$ and $\mathbf{I}$ is the identity tensor.
In two dimensions, the only component of $\boldsymbol{\omega}_o$ relevant to the dynamics of $\mathbf{q}_i$ will be that of the out-of-plane direction, which we describe with unit vector $\mathbf{e}_z$.
We can express $\boldsymbol{\omega}_o = \omega_o\mathbf{e}_z$, with the sign of $\omega_o$ determining the handedness of the chiral dynamics. 
When the pairwise forces $\mathbf{F}_{ij}$ are chosen such that particles effectively exclude area with diameter $d$, the system is fully characterized by three dimensionless quantities: the overall area fraction $\phi \equiv \rho \pi d^2/4$ (where $\rho$ is the number density); the dimensionlesss intrinsic run length, $\ell_o/d = U_o\tau_R/d$; and the ratio of chiral and rotary diffusion timescales, $\chi\equiv \omega_o\tau_R$.
We note that this dimensionless measure of the chirality also has a clear geometric interpretation.
The intrinsic run length $\ell_o$ is a measure of the persistence length of an ideal and achiral active particle trajectory while $U_o/|\omega_o|$ is the radius of the trajectory of an ideal particle in the absence of rotational diffusion.
The ratio of these two length scales is simply $\ell_o |\omega_o| /U_o \equiv |\chi|$.
We recover the achiral ABP model in the limit of $\chi \rightarrow 0$.

\textit{Phase Coexistence of Chiral Active Matter.--}
Understanding the bulk phase coexistence of a system is prerequisite to understanding its interface.
Here, we formulate a nonequilibrium theory for the phase diagram of cABPs following a recently proposed mechanical framework~\cite{Omar2023b}. 
As explicitly detailed in the Supporting Information (SI)~\cite{Note1}, we derive the dynamics of the density field, the relevant order parameter for fluid-fluid coexistence, beginning from the $N$-body distribution of microstates and Eq.~\eqref{eq:eom} (see Appendix~\ref{appendix:fokker-planck}).
Just as for ABPs, the microscopically derived mechanical balance for cABPs consists of the conservative stress ($\boldsymbol{\sigma}^{C}$) generated by particle interactions and an active stress ($\boldsymbol{\sigma}_{\rm act}$) induced by the active force with $\mathbf{0} = \boldsymbol{\nabla} \cdot (\boldsymbol{\sigma}^{C} + \boldsymbol{\sigma}_{\rm act})$.
For repuslive achiral ABPs, the active stress is the origin of the mechanical instability that underlies MIPS~\cite{Takatori2015TowardsMatter, Solon2015PressureSpheres}. 
Chiral dynamics directly alter the active stress by generating \textit{antisymmetric} active stresses (much like chiral active dumbbells~\cite{Klymko2017StatisticalHydrodynamics}) for finite $\chi$ while also diminishing the symmetric components.

While the total dynamic stress tensor $\boldsymbol{\Sigma} \equiv \boldsymbol{\sigma}_{\rm act} + \boldsymbol{\sigma}^{C}$ is no longer symmetric for finite $\chi$ (see Appendix~\ref{appendix:flux} for our derived expressions for $\boldsymbol{\Sigma}$) this proves to be inconsequential for macroscopic phase coexistence.
We consider two macroscopic coexisting phases with an interface which (on average) has a normal direction parallel to $\mathbf{e}_x$ and translational invariance in the $y$-direction.
In this scenario, only the symmetric contribution to the stress appears in the static mechanical balance which reduces to $0 = d\Sigma_{xx}/dx$.
We can then invoke the recently derived mechanical coexistence criteria~\cite{Omar2023b}:
\begin{subequations}
    \label{eq:coexist}
    \begin{equation}
        \mathcal{P}(\rho^{\rm liq}) = \mathcal{P}(\rho^{\rm gas}) \equiv \mathcal{P}^{\rm coexist},
    \end{equation}
    \begin{equation}
    \label{eq:weighted-area}
        \int_{\rho^{\rm liq}}^{\rho^{\rm gas}} \left[\mathcal{P} - \mathcal{P}^{\rm coexist}\right]E(\rho)\text{d}\rho,
    \end{equation}
where $\mathcal{P}(\rho)$ is the bulk contribution to the dynamic pressure, and $E(\rho)$ is determined from the interfacial contributions to the dynamic pressure. 
Equation~\eqref{eq:weighted-area} takes the form of a ``weighted-area'' Maxwell construction~\cite{Aifantis1983TheRule,Solon2018GeneralizedMatter,Omar2023b}. 

With a precise form for $\mathcal{P}(\rho)$ and $E(\rho)$, we can use Eq.~\eqref{eq:coexist} to determine the coexisting densities, $\rho^{\rm liq}$ and $\rho^{\rm gas}$. 
For cABPs we find (see Section~1.4 in the SI~\cite{Note1}):
    \begin{equation}
        \mathcal{P}(\rho) = p_C(\rho) + p_{\rm act}(\rho),
    \end{equation}
    \begin{equation}
    \label{eq:pact}
        p_{\rm act}(\rho) = \frac{\zeta U_o \ell_o \rho\overline{U}(\rho)}{2(1 + \chi^2)},
    \end{equation}
    \begin{equation}
        E(\rho) \sim \left[\overline{U}(\rho)\right]^{-\frac{\chi^2}{4}}\Biggr(\frac{\partial p_C}{\partial \rho}\Biggl)^{1 - \frac{\chi^2}{4}},
    \end{equation}
\end{subequations}
where $p_C(\rho)$ and $p_{\rm act}(\rho)$ are the bulk contributions to the conservative and active stresses, respectively. 
$\overline{U}(\rho)$ is the dimensionless average active speed which ranges from $0$ (interactions completely arrest particle motion) to $1$ (particles move unimpeded with speed $U_o$) which depends on the particle interactions and density.
Both the dynamic pressure and weighted-area construction variable reduce to those recently found for ABPs~\cite{Omar2023b} in the limit $\chi\rightarrow 0$, as anticipated.
To determine the qualitative impact of chirality, we adopt forms of the equations of state that have either been previously proposed ($p_C$) or are similar to established functions ($\overline{U}$) for repuslive ABPs (the precise forms are provided in Appendix~\ref{appendix:eqofstate}).

\begin{figure}
	\centering
	\includegraphics[width=0.48\textwidth]{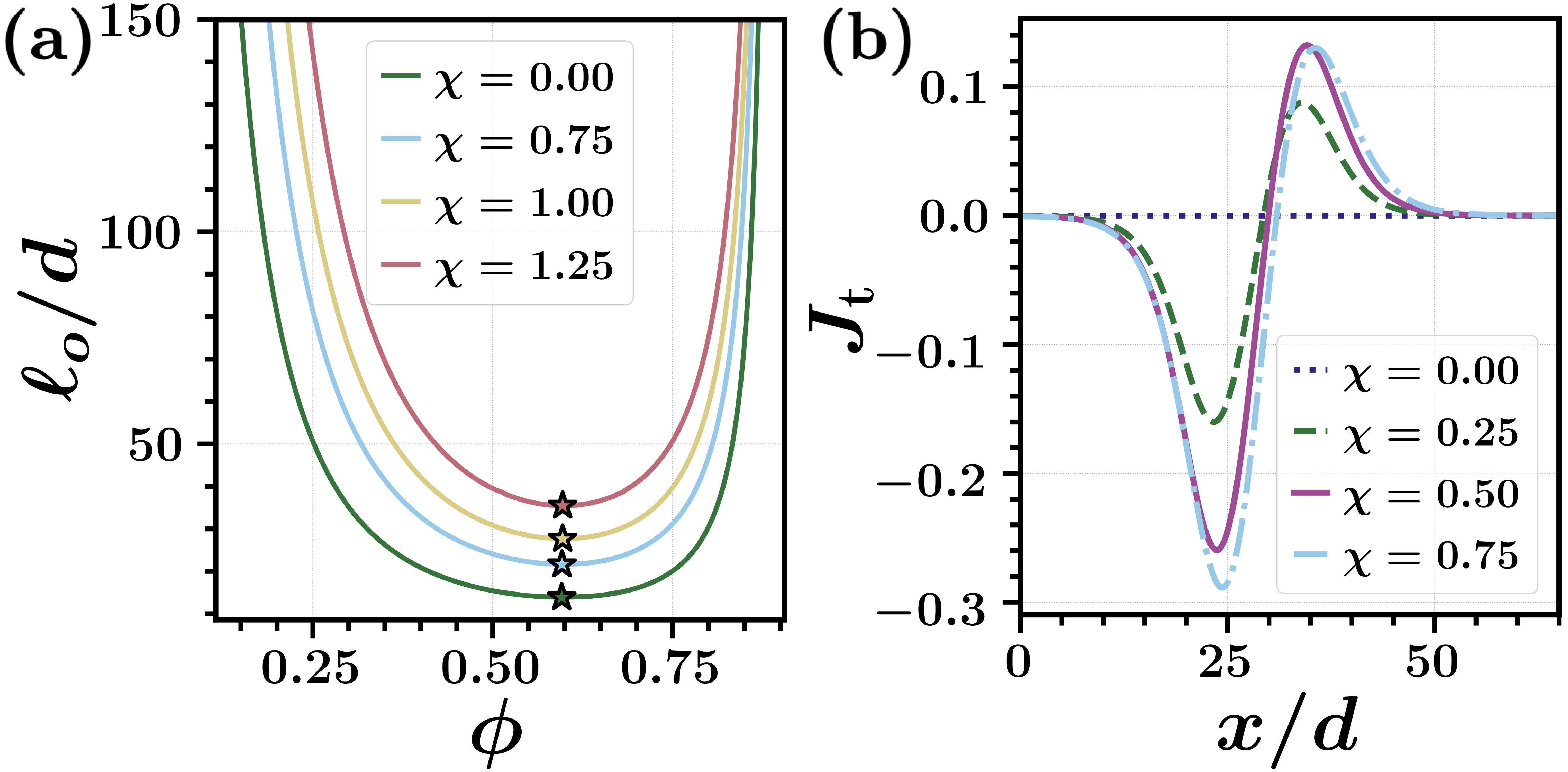}
	\caption{\protect\small{{(a) Theoretical phase diagram of chiral ABPs for selected values of $\chi$. Stars indicate critical points. (b) Theoretical steady-state $J_t(x)$ (units of $U_o/a_p$ where $a_p = \pi d^2 /4$ is the particle area) for $\ell_o/d = 125$. $J_t(x) \to 0$ as $\chi\to 0$.}}}
	\label{fig:figure1}
\end{figure}

The phase diagram for cABPs is shown in Fig.~\ref{fig:figure1}(a) as a function of activity for several values of  $\chi$.
Increasing chirality monotonically raises the critical activity indicating that \textit{chirality acts to stabilize active fluids}. 
This can be immediately understood by examining the form for the bulk dynamic pressure. 
For achiral ABPs, the nonmonontic dependence of the active pressure is what gives rise to MIPS. 
Chirality acts to reduce the impact of the active pressure through the $(1+\chi^2)$ factor present in Eq.~\eqref{eq:pact}.
Physically, the active pressure can be thought of as a measure of the \textit{effective persistence length} of the particle trajectories with $p_{\rm act} \propto \rho \zeta U_o \ell^{\rm eff}$ where $\ell^{\rm eff} \equiv U_o\overline{U}\tau_{R}/(1+\chi^2)$.
While the reduction in the effective run length arising from steric interactions has a significant density dependence (crucial for MIPS and reflected in $\overline{U}$), increasing chirality \textit{uniformly} reduces $\ell^{\rm eff}$ \textit{for all densities}.
Increasing $\chi$ causes particles to depart from their persistent trajectories and instead move in increasingly tight orbits, diminishing the active pressure. 
This quadratic rescaling of the effective run length with chirality is consistent with the analytical model of cABPs proposed in Ref.~\cite{Bickmann2022AnalyticalSwimmers}.
Finally, we note that the persistence motion of ABPs is what gives rise to many of their intriguing nonequilibrium behavior.
In addition to MIPS, these behaviors include accumulation at boundaries~\cite{Yan2015a} and inducing \textit{repulsive} depletion forces~\cite{Omar2019SwimmingDoping} between passive particles. 
The reduction in effective run length with chirality may be a useful perspective towards understanding why these behaviors can be altered through chirality~\cite{Batton2024MicroscopicEnvironments}.

Intriguingly, while the average density flux normal to the interface must vanish in this stationary coexistence scenario the \textit{tangential flux} can remain finite.
This stands in contrast to passive systems and active systems with achiral dynamics. 
While the flux associated with this density profile is zero in the direction normal to the interface, the flux \textit{tangential} to the interface (i.e.,~parallel to $\mathbf{e}_y$) may be nonzero when $\chi$ is finite. 
We can compute the spatial tangential number density flux $J_t(x)\equiv \mathbf{J}\cdot\mathbf{e}_y$ by first numerically determining the complete spatial density profile, $\upvarphi(x)$~\cite{Langford2023}.
We indeed find \textit{odd tangential flows} in the vicinity of the interface [see Fig.~\ref{fig:figure1}(b)] and we thus expect such flows to affect the interfacial dynamics, which existing theories of active interfaces~\cite{Fausti2021CapillarySeparation,Langford2023} have yet to account for. 
It is important to note that these finite tangential flux are permitted in our theory, but in an experiment which enforces no flux boundary conditions in all directions, the translational invariance of the density profile will be lost in the tangential direction, and details of the experimental geometry may become crucial for understanding phase behavior.

\textit{Fluctuating Hydrodynamics of Chiral Active Matter.--}
To construct a theory for chiral interfacial dynamics, we first build a fluctuating hydrodynamic~\cite{Dean1996LangevinProcesses,Cugliandolo2015StochasticDipoles} description of cABPs~\cite{Note1}:
\begin{subequations}
 \label{eq:flucthydro}   
\begin{equation}
    \frac{\partial\rho}{\partial t} = -\boldsymbol{\nabla}\cdot\mathbf{J} ,
\end{equation}
\begin{equation}
    \mathbf{J} = \frac{1}{\zeta}\boldsymbol{\nabla}\cdot\bm{\Sigma} + \bm{\eta}^{\rm act},\label{eq:flux}
\end{equation}
\end{subequations}
where $\bm{\eta}^{\rm act}$ is a stochastic component of the flux with zero mean and correlations:
\begin{align}
    \langle \bm{\eta}^{\rm act}(\mathbf{r},t)\bm{\eta}^{\rm act}(\mathbf{r}',t') \rangle = &2\frac{k_BT^{\rm act}}{\zeta}\left(\rho\mathbf{I}-2\mathbf{Q}\right)\nonumber \\ &\times\delta(\mathbf{r}'-\mathbf{r})\delta(t-t').
\end{align}
We note that the derivation of these dynamics has assumed the magnitude of the flux $|\mathbf{J}|$ to be small.
Here, $k_BT^{\rm act} \equiv \zeta U_o \ell_o / 2$ is the athermal active energy scale and $\mathbf{Q}$ is the traceless nematic order tensor, which is a closed function of $\rho$ within our approximations.
Full expressions for the stress tensor contributions are provided in Appendix~\ref{appendix:flux}.
It is worth noting that the antisymmetric stress will affect the flux but cannot affect the density field dynamics as two divergences of any antisymmetric tensor must vanish.
However, spatial parity violating terms (e.g.,~terms proportional to $\partial_x\rho \partial_y\rho$) also arise in the symmetric stress which \textit{does} impact the density field dynamics (see Eq.~\ref{aeq:dynstress}).

\textit{Capillary Fluctuations and Odd Surface Flows.--}
Now that we have described the flux-free phase separated states \textit{and} weak fluctuations of the stochastic density field, we are now well-poised to probe dynamics of the interface. 
In order to connect the fluctuations of the density field to that of the interface, we introduce the linearizing ansatz proposed by Bray \textit{et al.}~\cite{Bray1994TheoryKinetics,Bray2002InterfaceShear}:
\begin{equation}
    \rho(\mathbf{r},t) = \upvarphi[x - h(y)],\label{eq:ansatz}
\end{equation}
which assumes $|\partial_y h|<<1$.

As demonstrated by Ref.~\cite{Fausti2021CapillarySeparation} and recently applied to microscopic systems~\cite{Langford2023}, substitution of Eq.~\eqref{eq:ansatz} into Eqs.~\eqref{eq:flucthydro} and integration after multiplication by the appropriate Green's function as well as the appropriate pseudovariable~\footnote{Note2} results in a linearized Langevin equation for the interfacial dynamics.
This Fourier transformed Langevin equation with $k$ corresponding to a wavevector parallel to $\mathbf{e}_y$ (see SI~\cite{Note1} and Ref.~\cite{Langford2023} for details) is given by:
\begin{equation}
    \zeta_{\rm eff}\frac{\partial h}{\partial t} = -k^2|k|\upgamma_{\rm cw}h - ik|k|\nu_{\rm odd}h + \xi^{\rm iso} + \xi^{\rm aniso} + \mathcal{O}\left(h^2k^3\right)\label{eq:interfacelangevin},
\end{equation}
where $\zeta_{\rm eff}$ is an effective drag, $\upgamma_{\rm cw}$ is the capillary tension~\cite{Fausti2021CapillarySeparation,Langford2023}, and we have introduced $\nu_{\rm odd}$, the \textit{odd flow coefficient}.
$\xi^{\rm iso}$ is a noise originating from the isotropic components of $\bm{\eta}^{\rm act}$ with zero average and correlations:
\begin{equation}
    \langle \xi^{\rm iso}(k,t)\xi^{\rm iso}(k',t') \rangle = 4\pi|k|(k_BT)^{\rm act}\zeta_{\rm eff}\delta(k+k')\delta(t-t').
\end{equation}
Full expressions for $\upgamma_{\rm cw}$, $\nu_{\rm odd}$, $\zeta_{\rm eff}$, and the correlations of $\xi^{\rm aniso}$ are provided in Appendix~\ref{appendix:interface}.

Equation~\eqref{eq:interfacelangevin} differs from those derived by Refs.~\cite{Fausti2021CapillarySeparation,Langford2023} due to the complex term proportional to $\nu_{\rm odd}$, which couples real and imaginary parts of $h(k)$, thus generating flows determined by the sign of $\chi$. 
However, the criteria for a stable interface remains $\upgamma_{\rm cw} > 0$~\cite{Note1}.
In this long wavelength limit (where $\xi^{\rm aniso}$ may be safely discarded~\cite{Langford2023}), the stationary fluctuations of $h(k)$ follow the familiar scaling of capillary-wave theory with~\cite{Note1}:
\begin{equation}
    \langle |h(k)|^2\rangle = \frac{k_BT^{\rm act} L}{\upgamma_{\rm cw}k^2},\label{eq:interfaceflucts}
\end{equation}
where $L$ is the length of the system in the $y$-dimension.
Additionally, in this long wavelength limit, the distribution of capillary waves may be determined from the steady-state Fokker-Planck equation~\cite{Zwanzig2001NonequilibriumMechanics,Note1} corresponding to Eq.~\eqref{eq:interfacelangevin}:
\begin{equation}
    P[h(k)] \sim \exp{\left[-\frac{k^2\upgamma_{\rm cw}\left(\text{Re}\bigl(h(k)\bigr)^2 + \text{Im}\bigl(h(k)\bigr)^2\right)}{2L k_BT^{\rm act}}\right]}.\label{eq:interfaceboltzmann}
\end{equation}
Surprisingly, only the capillary tension $\upgamma_{\rm cw}$ enters the stationary statistics of $h(k)$. 
From Eqs.~\eqref{eq:interfaceflucts} and~\eqref{eq:interfaceboltzmann}, one can see that interfacial fluctuations of chiral active interfaces are well-described by a surface-area minimizing Boltzmann distribution proportional to the interfacial stiffness $\upgamma_{\rm cw}/k_BT^{\rm act}$ just as their achiral counterparts~\cite{Fausti2021CapillarySeparation,Langford2023}.
The odd flow coefficient $\nu_{\rm odd}$, while crucial for describing the travelling-wave \textit{dynamics} of $h(k)$, is irrelevant to the steady-state distribution governing the \textit{structure} of $h(k)$.
However, the impact of $\nu_{\rm odd}$ is evident in the power spectrum which (for low $k$) takes the form:
\begin{equation}
    \langle |h(k,\omega)|^2 \rangle = \frac{2|k|k_BT^{\rm act}\zeta_{\rm eff}L\Gamma }{|k|^6\upgamma_{\rm cw}^2 + \left(\omega\zeta_{\rm eff} + k|k|\nu_{\rm odd}\right)^2},\label{eq:powerspec}
\end{equation}
where $\Gamma$ is the total time duration of the observation window.  

\begin{figure}
	\centering
	\includegraphics[width=0.48\textwidth]{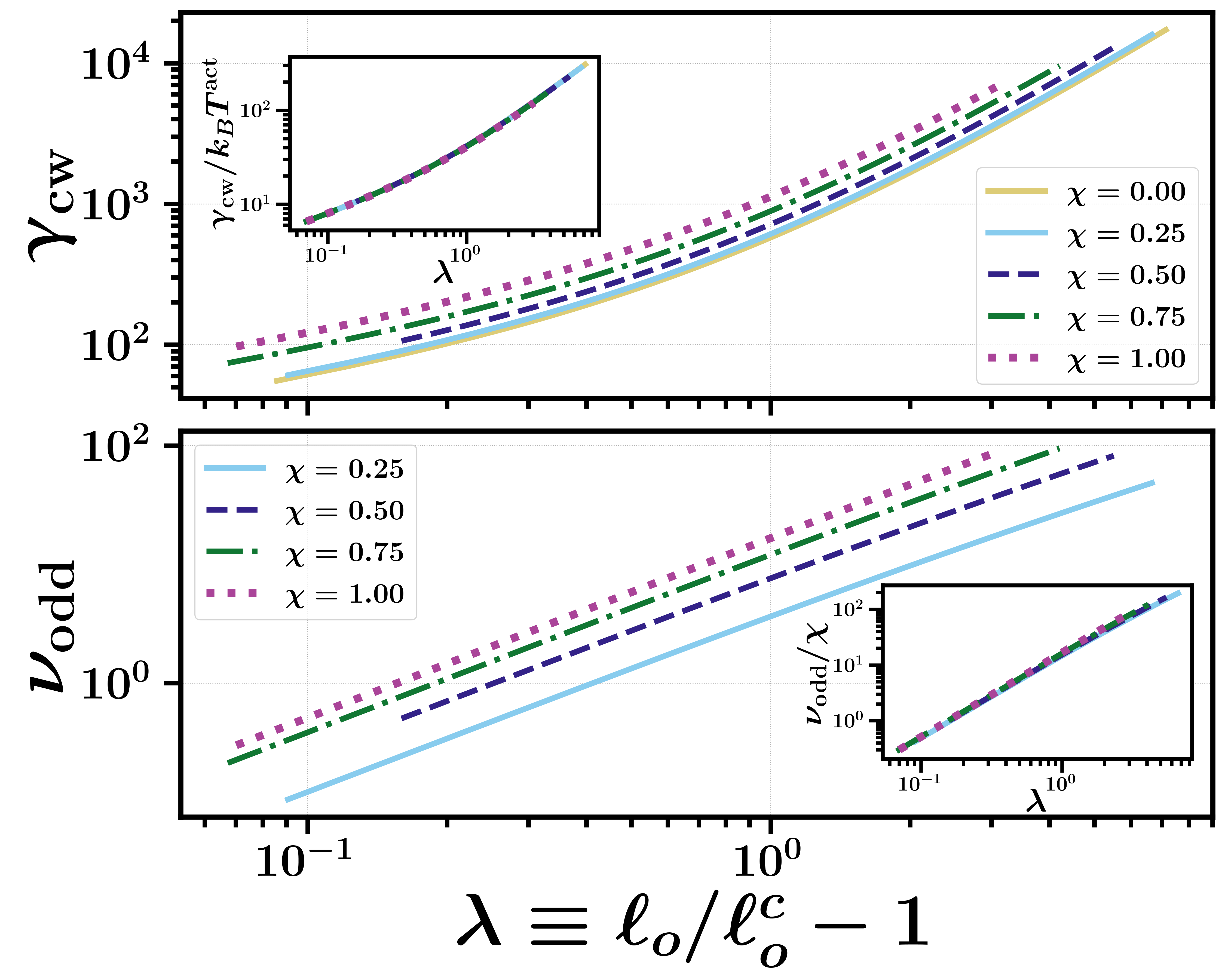}
	\caption{\protect\small{{$k\to 0$ limit of $\upgamma_{\rm cw}$ (units of $\zeta U_o/d$) and $\nu_{\rm odd}$ (units of $\zeta U_o/d^2$) as a function of reduced run length. Insets demonstrate collapse of curves from normalization of $\upgamma_{\rm cw}$ and $\nu_{\rm odd}$ by $k_BT^{\rm act}$ and $\chi$, respectively.  }}}
	\label{fig:figure2}
\end{figure}

Inspecting the full expression for $\upgamma_{\rm cw}$[see Eq.~\eqref{appeq:gammacw}], we find that its form is similar to that of parity-symmetric active systems and therefore should obey similar scaling.
The primary influence of $\chi$ on the interfacial stiffness is through its impact on the critical point. 
We therefore compute $\upgamma_{\rm cw}(k\to 0)$ as a function of a \textit{reduced} run length $\lambda \equiv \ell_o/\ell_o^c - 1$ [see Fig.~\ref{fig:figure2}], where the critical activity $\ell_o^c$ now increases with $\chi$.
Guided by our theory, a complete curve collapse is achieved upon normalization by $k_BT^{\rm act}$ as shown in the Fig.~\ref{fig:figure2} inset.
Similarly, inspecting the form of $\nu_{\rm odd}$ [see Eq.~\eqref{appeq:nuodd}] we find that it is linearly proportional to $\chi$. 
Evaluating $\nu_{\rm odd}(k\rightarrow 0)$ as a function of $\lambda$ [included in Fig.~\ref{fig:figure2}] also reveals a complete collapse of the odd flow coefficient upon normalization by $\chi$.

\begin{figure}
	\centering
	\includegraphics[width=0.48\textwidth]{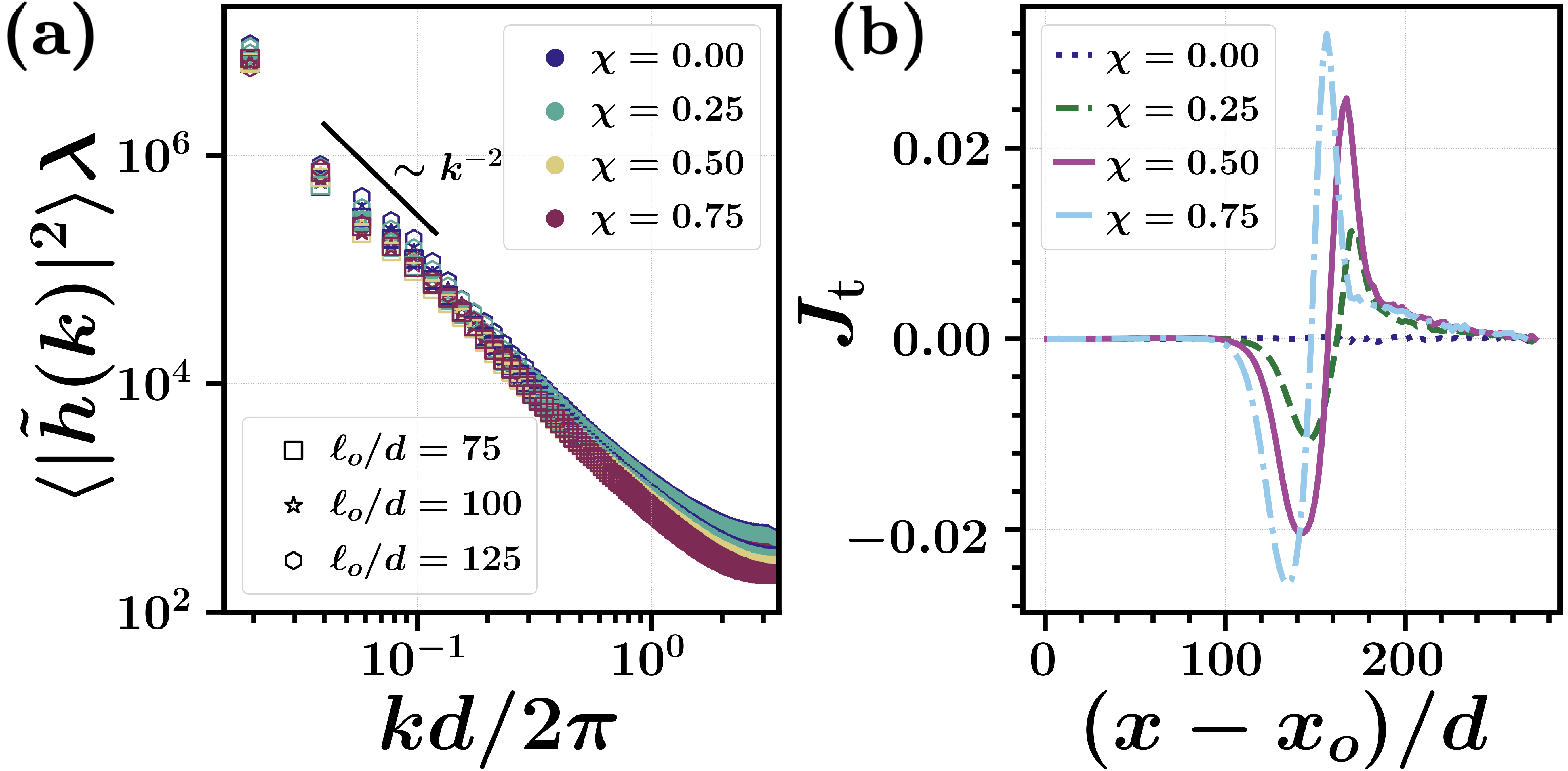}
	\caption{\protect\small{{ (a) Stationary fluctuation spectra $\langle |\tilde{h}(k)|^2\rangle\lambda$ (units of $d^4$) for twelve combinations of $\ell_o/d$ and $\chi$ from BD simulations. $k^{-2}$ scaling is measured across over a decade of $k$ values, with error bars reported in the SI~\cite{Note1}. Predicted collapse from multiplication by $\lambda$ is also observed. (b) Steady-state $J_t(x)$ (units of $U_o/a_p$) from BD simulations for $\ell_o/d = 125$.}}}
	\label{fig:figure3}
\end{figure}

\textit{Particle Simulation Data.--}
Our theory predicts that the breaking of spatial-parity symmetry ($\chi \neq 0$) results in odd surface flows at the MIPS interface and that these odd surface flows play \textit{no role} in the stationary statistics of $h(k)$, which is entirely controlled by the interfacial stiffness $\upgamma_{\rm cw}/k_BT^{\rm act}$.
To test these predictions, we conduct large-scale Brownian dynamics (BD) simulations~\cite{Anderson2020} of macroscopically phase separated cABPs at an array of values of $\chi$ and $\ell_o/d$, with simulation details provided in Appendix~\ref{appendix:simulation}. 
The instantaneous shape of the interface $h(y)$ is extracted directly from Brownian dynamics~\cite{Patch2018}.
The stationary height fluctuations  $\langle |\tilde{h}(k)|^2 \rangle$ are then computed from $\tilde{h}(k)$, the \textit{discrete} Fourier transform of $h(y)$.
Figure~\ref{fig:figure2} reveals both a collapse in the interfacial stiffness when plotted as a function of $\lambda$ \textit{and} a linear scaling with $\lambda$ at high run length length.
This theoretical prediction, combined with Eq.~\eqref{eq:interfaceflucts}, implies that the product of $\langle |\tilde{h}(k)| \rangle$ and $\lambda$ should collapse across all values of $\ell_o$ and $\chi$.
The agreement between our theoretical prediction and particle-based simulations is excellent [see Fig.~\ref{fig:figure3}(a)], demonstrating that chirality only impacts the stationary statistics by altering the MIPS critical point.
The predicted $k^{-2}$ is also robustly measured with statistical uncertainty reported in the SI~\cite{Note1}.

The odd flow coefficient $\nu_{\rm odd}$ may be inferred from the power spectrum [see Eq.~\eqref{eq:powerspec}], but in practice the measurement is challenging statistically. 
We therefore directly measure the odd surface flows by recording the tangential flux of particles, $J_t(x)$, as theoretically predicted in Fig.~\ref{fig:figure1}(b).
$\nu_{\rm odd}$ is finite when $\chi \neq 0$ and Eq.~\eqref{eq:interfacelangevin} implies that a net velocity of particles will develop in the vicinity of the interface.
Our simulations corroborate this while also revealing that the spatial profile of $J_t(x)$ [see Fig.~\ref{fig:figure3}(b)] is qualitatively similar to that predicted by our theory [see Fig.~\ref{fig:figure1}(b)]. 
The absence of quantitative agreement likely stems from a combination of factors such as the employed equations of state~\cite{Note1} and the broadening of the interface due to capillary-waves~\cite{Evans1979TheFluids,Rowlinson1982MolecularCapillarity} not accounted for in the theoretically predicted profiles.
Nonetheless, our particle-based simulations still verify all qualitative trends predicted by our theory.

\textit{Conclusions.-- }
We derive a nonequilibrium theory of phase coexistence, interfacial fluctuations, and odd surface flows for a model system of chiral active matter.
We find that chirality both reduces the effective motility and induces non-zero currents tangential to the interface at steady-state, the latter of which emerges in the form of an imaginary term proportional to an odd flow coefficient $\nu_{\rm odd}$ in the interfacial dynamics.
Despite the intriguing dynamics arising from these odd flows, the stability of interfacial dynamics and recovery of a surface-area minimizing principle remains controlled by $\upgamma_{\rm cw}$.
Large-scale Brownian dynamics simulations corroborate our qualitative theoretical predictions.
By extending the nonequilibrium theory of phase coexistence~\cite{Aifantis1983TheRule,Solon2018GeneralizedMatter,Omar2023b} and (linear) interfacial dynamics~\cite{Bray2002InterfaceShear,Fausti2021CapillarySeparation,Langford2023} to parity-breaking systems, it is our hope that these perspectives now extend to experimentally realizable active matter systems. 
Finally, we note that Ma \textit{et al.}~\cite{Ma2022DynamicalParticles} recently demonstrated that cABPs can exhibit a diffusive instability that results in ``dynamic clusters'' which can preclude the observation of MIPS.
A complete stability/phase diagram for chiral active matter, which includes both MIPS and the dynamic clustering instability, remains an outstanding challenge that is the subject of future work.

\begin{acknowledgments}
\textit{Acknowledgments.-- }
This material is based upon work supported by the U.S. Department of Energy, Office of Science, under Award Number DE-SC0024900. 
L.L. was supported in part by the Department of Defense (DoD) through the National Defense Science \& Engineering Graduate (NDSEG) Fellowship Program. 
\end{acknowledgments}

\newpage
\section{End Matter}

\fakesection{Appendix I: Fokker-Planck and Continuity Equations}\label{appendix:fokker-planck}
\textit{Appendix I: Fokker-Planck and Continuity Equations.--}
The Fokker-Planck equation associated with Eqs.~\eqref{eq:eom} is given by:
\begin{equation}
    \label{seq:fokkerplanck}
    \frac{\partial}{\partial t}f(\bm{\Gamma};t)= \mathcal{L}f(\bm{\Gamma};t),\tag{Ia}
\end{equation}
where $\bm{\Gamma}\equiv [\mathbf{r}_i^N,\mathbf{q}_i^N]$ is a $4N$ dimensional vector and $\mathcal{L}$ is the dynamical operator consistent with Eqs.~\eqref{eq:eom} with:
\begin{equation}
    \mathcal{L} \equiv \sum_{i=1}^{N}\left[\frac{\partial}{\partial \mathbf{r}_i}\cdot\left(-U_o\mathbf{q}_i - \frac{1}{\zeta}\sum_{j\neq i}^{N}\mathbf{F}_{ij}\right) + \boldsymbol{\nabla}_i^{R}\cdot\left( D_R\boldsymbol{\nabla}_i^R -\bm{\omega}_o \right)\right].\tag{Ib}
\end{equation}
Here, we have defined the rotational gradient operator as $\boldsymbol{\nabla}_i^{R} \equiv \mathbf{q}_i\times \partial /\partial \mathbf{q}_i$.
The adjoint, $\mathcal{L}^*$, of $\mathcal{L}$ is provided in the SI~\cite{Note1}.
The average of an arbitrary macroscopic observable $\mathcal{O}$ is $\langle \mathcal{O} \rangle = \int\text{d}\bm{\Gamma}\mathcal{O} f(\bm{\Gamma};t) $, while the average evolution $\mathcal{O}$ can be expressed as:
\begin{equation}
    \frac{\partial \langle \mathcal{O} \rangle}{\partial t} = \int \text{d}\bm{\Gamma}  f(\bm{\Gamma};t)\mathcal{L}^*\mathcal{O}.\tag{Id}
\end{equation}
We now define the average number density ${\rho \equiv \langle \sum_{i=1}^{N}\delta(\mathbf{r}_i-\mathbf{r}) \rangle}$.
Then, the evolution of the density is found to be~\cite{Note1}:
\begin{equation}
\label{appeq:denseevolve}
    \frac{\partial \rho}{\partial t} = -\boldsymbol{\nabla}\cdot\mathbf{J},\tag{Ie}
\end{equation}
where the number density flux, $\mathbf{J}$, is:
\begin{equation}
    \mathbf{J} = U_o \mathbf{m}  + \frac{1}{\zeta}\boldsymbol{\nabla}\cdot\bm{\sigma}^C.\tag{If}
\end{equation}
Here, $\boldsymbol{\nabla} \equiv \partial/\partial \mathbf{r}$ and we have defined the polar order density $\mathbf{m} \equiv \left\langle \sum_{i=1}^{N}\mathbf{q}_i\delta(\mathbf{r}_i-\mathbf{r})\right\rangle$ and interaction stress $\bm{\sigma}^C$ (see Refs.~\cite{Omar2023b, Langford2023} for microscopic expression).
The evolution of the density is dependent on the polar order is found to be~\cite{Note1}:
\begin{subequations}
\label{appq:polarevolve}
    \begin{equation}
        \frac{\partial\mathbf{m}}{\partial t} = -\boldsymbol{\nabla}\cdot\mathbf{J}^{m} -D_R\mathbf{m}  + \bm{\omega}_o\times\mathbf{m},\tag{Ig}
    \end{equation}
where we have defined a flux of polar order $\mathbf{J}^{\mathbf{m}}$ as:
    \begin{equation}
        \mathbf{J}^{m} = U_o\left(\mathbf{Q} + \frac{\rho}{d}\mathbf{I}\right) + \frac{1}{\zeta}\bm{\kappa^m} + \frac{1}{\zeta}\boldsymbol{\nabla}\cdot\bm{\Sigma^m}.\tag{Ih}
    \end{equation}
\end{subequations}
Here $\bm{\kappa^m}$, $\bm{\Sigma^m}$, are ``stress'' and ``body force''-like contributions (see Refs.~\cite{Omar2023b, Langford2023}) and we have defined the traceless nematic order tensor as $\mathbf{Q} \equiv \left\langle \sum_{i=1}^{N} \left(\mathbf{q}_i\mathbf{q}_i - \frac{1}{2}\mathbf{I}\right)\delta(\mathbf{r}_i-\mathbf{r}) \right\rangle$.
The evolution of the polar order is dependent on the traceless nematic order which evolves according to:
\begin{subequations}
\label{seq:tracelessnematicevolve}
    \begin{equation}
        \frac{\partial\mathbf{Q}}{\partial t} = -\boldsymbol{\nabla}\cdot\mathbf{J}^Q  + \omega_o\begin{bmatrix}2Q_{xy} & Q_{xx}-Q_{yy} \\ Q_{xx}-Q_{yy} & 2Q_{xy}\end{bmatrix}- 4D_R\mathbf{Q},\tag{Ii}
    \end{equation}
where we have defined a flux of nematic order $\mathbf{J}^{Q}$ as:
    \begin{equation}
        \mathbf{J}^{Q} = U_o\tilde{\mathbf{B}} + \frac{1}{\zeta}\bm{\kappa^Q} + \frac{1}{\zeta}\boldsymbol{\nabla}\cdot\bm{\Sigma^Q} - \frac{U_o}{2}\mathbf{m}\mathbf{I} - \frac{1}{2\zeta} \left(\boldsymbol{\nabla}\cdot\bm{\sigma}^C\right)\mathbf{I}.\tag{Ij}
    \end{equation}
\end{subequations}
Here $\bm{\kappa^Q}$, $\bm{\Sigma^Q}$, are ``stress'' and ``body force''-like contributions  (see Refs.~\cite{Omar2023b, Langford2023}) and we have defined $\tilde{\mathbf{B}} \equiv \left\langle \sum_{i=1}^{N} \mathbf{q}_i\mathbf{q}_i\mathbf{q}_i \delta(\mathbf{r}_i-\mathbf{r}) \right\rangle$.
We find that the evolution of the traceless nematic order is dependent on $\tilde{\mathbf{B}}$. 
The coexistence theory proposed in Ref.~\cite{Omar2023b} requires a second order expansion of the dynamic stress tensor, or equivalently a third order expansion of the density flux.
Because the dynamics of the $(n)$th orientational moment is dependent on the divergence of the $(n+1)$th orientational moment, each additional level included in the hierarchy of equation serves to increase the spatial order.
Truncation of the hierarchy of equations at the nematic order (and approximating $\mathbf{B}$ as isotropic) is then sufficient for understanding coexistence.

\fakesection{Appendix II: Flux Components }\label{appendix:flux}
\textit{Appendix II: Dynamic Stress.--}
We take the stationary limit of the conservation equations, neglect the effects of  $\bm{\Sigma}^{\bm{m}/\bm{Q}}$ (justified theoretically and numerically in Ref.~\cite{Omar2023b}), and propose constitutive relations connecting $\bm{\kappa}^{\bm{m}/\bm{Q}}$ to the active speed $\overline{U}(\rho)$~\cite{Note1}.
In addition, following Refs.~\cite{Solon2018GeneralizedEnsembles,Omar2023b,Evans2023}, we neglect gradient contributions to the interaction stress and approximate $\bm{\sigma}^C \approx -p_C(\rho)\mathbf{I}$.
These approximations allow the density flux to be expressed as the divergence of a dynamic stress tensor $\mathbf{J} = \boldsymbol{\nabla}\cdot\bm{\Sigma}/\zeta$, which can further be broken into:
\begin{equation}
    \boldsymbol{\Sigma} = \mathbf{T} + \mathbf{T}^{\rm cross} + \mathbf{T}^{\rm odd}.\tag{IIa}
\end{equation}
$\mathbf{T}$ is found to be:
\begin{subequations}
\begin{align}
\label{aeq:dynstress}
    \mathbf{T} = \biggl[&-\mathcal{P}(\rho) + a(\rho)\boldsymbol{\nabla}^2\rho + \frac{\chi}{2}b(\rho)\frac{\partial\rho}{\partial x}\frac{\partial \rho}{\partial y} \nonumber \\ &+\frac{\chi^2}{8}b(\rho)\left[\left(\frac{\partial \rho}{\partial x}\right)^2-\left(\frac{\partial \rho}{\partial y}\right)^2\right]\biggr]\mathbf{I}  + b(\rho)\boldsymbol{\nabla}\rho\boldsymbol{\nabla}\rho \nonumber \\ &+ \frac{\chi}{4}b(\rho)\left[\left(\frac{\partial \rho}{\partial x}\right)^2-\left(\frac{\partial \rho}{\partial y}\right)^2\right]\left(\mathbf{e}_x\mathbf{e}_y + \mathbf{e}_y\mathbf{e}_x\right),\tag{IIb}
\end{align}
where $a(\rho)$ and $b(\rho)$ are defined as
\begin{equation}
    a(\rho) = \frac{3\ell_o^2}{16(1+\chi^2)}\overline{U}^2\frac{\partial p_C}{\partial \rho},\tag{IIc}
\end{equation}
and
\begin{equation}
    b(\rho) = \frac{3\ell_o^2}{16(1+\chi^2)}\overline{U}\frac{\partial}{\partial \rho}\left[\overline{U}\frac{\partial p_C}{\partial \rho}\right].\tag{IId}
\end{equation}
$\mathbf{T}^{\rm cross}$ can be expressed compactly as:
\begin{equation}
    \mathbf{T}^{\rm cross} = T_{xy}\left(\mathbf{e}_y\mathbf{e}_y - \mathbf{e}_x\mathbf{e}_x\right) + \chi\left( T_{xx} + \mathcal{P}\right)\left(\mathbf{e}_x\mathbf{e}_y + \mathbf{e}_y\mathbf{e}_x\right),\tag{IIe}
\end{equation}
while $\mathbf{T}^{\rm odd}$ is given by:
\begin{equation}
    \mathbf{T}^{\rm odd} = \begin{bmatrix}0 & -\frac{\chi\zeta U_o \ell_o\overline{U}\rho}{2(1+\chi^2)} \\ \frac{\chi\zeta U_o \ell_o\overline{U}\rho}{2(1+\chi^2)} & 0\end{bmatrix}.\tag{IIf}
\end{equation}
\end{subequations}

\fakesection{Appendix III: Equations of State }\label{appendix:eqofstate}
\textit{Appendix III: Equations of State.--}
Evaluation of the steady-state binodals, density profiles, and tangential fluxes requires equations of state (also provided in Section 1.6 of the SI~\cite{Note1}) for the interaction pressure $p_C(\rho)$ and the dimensionless active speed $\overline{U}$.
For simplicity, we choose equations of state that capture the expected qualitative behavior while satisfying known physical limits. 
For example, in the limit of zero area fraction we demand that the interaction pressure is zero and the dimensionless active speed is one. 
In the limit of close-packing ($\phi \rightarrow \phi_o$) we demand that the interaction pressure diverges and the effective active speed is zero.
We therefore adopt the interaction pressure proposed in Ref.~\cite{Takatori2015TowardsMatter}:
\begin{equation}
    \frac{p_C d}{\zeta U_o} = \frac{8}{\pi}\phi^2\left(1 - \frac{\phi}{\phi_o}\right)^{-1},\tag{IIIa}
\end{equation}
where $\phi \equiv \pi \rho d^2 /4$ is the area fraction and a value of $0.9$ is used for the close-packed area fraction $\phi_o$.
We use the following form for the effective active speed:
\begin{equation}
    \overline{U} = \left[1 + \phi\left(1 - \frac{\phi}{\phi_o}\right)^{-1}\right]^{-1}.\tag{IIIb}
\end{equation}

\fakesection{Appendix IV: Interfacial Quantities }\label{appendix:interface}
\textit{Appendix IV: Interfacial Quantities.--}
As derived in the SI~\cite{Note1}, the capillary tension $\upgamma_{\rm cw}$ has the following form:
\begin{align}
    &\upgamma_{\rm cw} = \frac{A(k)}{\rho^{\rm surf}B(k)} \Biggl[ \int dx \frac{\partial\mathcal{E}_{\rm cw}}{\partial x}\frac{\partial \upvarphi}{\partial x}a(\rho) - \left(1 - \frac{\chi^2}{2}\right)\nonumber \\ & \times \int dx \frac{\partial\mathcal{E}_{\rm cw}}{\partial x}\int dx' e^{-|k||x-x'|}\text{sgn}(x-x') b(\rho)\left(\frac{\partial \upvarphi}{\partial x'}\right)^2\Biggr],\tag{IVa}\label{appeq:gammacw}
\end{align}
where $\text{sgn}$ is a function which returns $+1$ ($-1$) given a positive (negative) argument and the odd flow coefficient $\nu_{\rm odd}$ was found to be:
\begin{align}
    & \nu_{\rm odd } = \frac{A(k)}{\rho^{\rm surf}B(k)}\Biggl[ \frac{\chi}{2}\int dx \frac{\partial\mathcal{E}_{\rm cw}}{\partial x} b(\rho)\left(\frac{\partial \upvarphi}{\partial x}\right)^2\nonumber \\ & +|k|\chi\int dx \frac{\partial\mathcal{E}_{\rm cw}}{\partial x}\int dx' e^{-|k||x-x'|}b(\rho)\left(\frac{\partial \upvarphi}{\partial x'}\right)^2 \nonumber \\ &- k\chi\int dx \frac{\partial\mathcal{E}_{\rm cw}}{\partial x}\int dx' e^{-|k||x-x'|} \text{sgn}(x-x')a(\rho)\frac{\partial \upvarphi}{\partial x'} \Biggr].\tag{IVb}\label{appeq:nuodd}
\end{align}
We have also defined an effective drag coefficient $\zeta_{\rm eff}$ as:
\begin{align}
    \zeta_{\rm eff} = \frac{\zeta A^2(k)}{2B(k) \rho^{\rm surf}},\tag{IVc}
\end{align}
and the covariance of the anisotropic fluctuations $\xi^{\rm aniso}$ is given by:
\begin{align}
    \langle \xi^{\rm aniso}(k,t)\xi^{\rm aniso}(k',t') \rangle = &\frac{\zeta_{\rm eff}\left(C(k) + D(k) + R(k)\right)}{\rho^{\rm surf}B(k)}\nonumber \\ &\times(2\pi)\delta(t-t')\delta(k+k').\tag{IVd}
\end{align}
$A(k)$, $B(k)$, $C(k)$, $D(k)$ and $R(k)$ are all functions which must be evaluated numerically and have forms detailed in Section 2.2 of the SI~\cite{Note1}.
In the above equations $\mathcal{E}_{\rm cw}$ is the capillary psueodvariable where $E_{\rm cw}(\rho) \equiv \partial_{\rho}\mathcal{E}_{\rm cw}$ and $E_{\rm cw}(\rho)$ is expressed as:
\begin{equation}
    E_{\rm cw} \sim \left(\overline{U}\right)^{\frac{\chi^2}{4}}\left(\frac{\partial p_C}{\partial \rho}\right)^{1 + \frac{\chi^2}{4}}.\tag{IVe}
\end{equation}
The fact $E_{\rm cw}(\rho)\neq E(\rho) $ for cABPs but not standard ABPs has clear physical interpretation.
The pseudovariable used to construct the binodals was chosen to eliminate nonlinear gradients in the $x$ direction while the pseudovariable used to derive the interfacial dynamics was chosen to eliminate nonlinear gradients in the $y$ direction.
Therefore, systems which violate spatial parity will generally have different pseudovariables used for deriving linear interfacial dynamics and binodals.

\fakesection{Appendix IV: Simulation Details }\label{appendix:simulation}
\textit{Appendix IV: Simulation Details.--}
Brownian dynamics simulations of over $10^5$ cABPs were conducted at each combination of the parameters $\chi \in \{0.00,0.25,0.50,0.75\}$ and $\ell_o/d \in \{75,100,125\}$. 
We choose $\mathbf{F}_{ij}$ to correspond to that of polydisperse hard disks~\cite{Weeks1971,Phillips2010StabilitySystems} with diameters set by a Gaussian distribution with average $d$ and standard deviation $0.1d$ such that no hexatic phases are present and $\rho$ can be taken as the sole order parameter of the phase separation. 
Hard-disk statistics were approximated by using a WCA potential~\cite{Phillips2010StabilitySystems} with a fixed stiffness of $\mathcal{S}\equiv \epsilon / \zeta U_o d_{\rm LJ} = 50$, where $\epsilon$ and $d_{\rm LJ}$ are the Lennard-Jones energy scale and (mean) diameter, respectively. 
The effective exclusion diameter is $d \equiv 2^{1/6} d_{\rm LJ}$.
Upon reaching a steady state of MIPS, we run simulations for an additional minimum duration of $6.6\times 10^{4} d/U_o$. 
Dynamical quantities were measured from simulations with durations of $2.4\times 10^{5}d/U_o$.
All simulations were executed using the \texttt{HOOMD-Blue} software package~\cite{Anderson2020}.

The instantaneous shape of the interface $h(y)$ was measured using the algorithm developed by Patch \textit{et al.}~\cite{Patch2018} every $4.45 d/U_o$. 
$\tilde{h}(k)$ was then calculated from the discrete Fourier transform of $h(y)$, and the average quantity $\langle |\tilde{h}(k)|^2 \rangle$ is reported in Fig.~\ref{fig:figure3}(a).
The average tangential flux profiles reported in Fig~\ref{fig:figure3}(b) were calculated by measuring the average density profiles and multiplying by the average velocity profile. 
We note that velocity is an ill-defined concept for particles undergoing Brownian dynamics, we therefore measured the $y$ displacement of particles over a finite period of time, i.e. $(\mathbf{r}_i(t_2) - \mathbf{r}_i(t_1))\cdot\mathbf{e}_y/(t_2-t_1)$, similar to the method employed by Ref.~\cite{DelJunco2019}. 
A constant time separation of $t_2 - t_1 = 4.45 d/U_o$ was chosen.
\end{document}


\maketitle

\pagenumbering{roman}

\setcounter{secnumdepth}{2}
\setcounter{tocdepth}{2}
{
	\hypersetup{
		linkcolor=Black,
		citecolor=Black
	}
	\vspace{-30pt}
	\tableofcontents
}

\noindent\rule{5.0cm}{0.4pt}
	{\footnotesize
	$\\$
	{
	    {$^\dag \,$}aomar@berkeley.edu}
	}

\newpage 
\pagenumbering{arabic}
\section{\label{sec:coexist}Coexistence of Chiral Active Particles}
In this Section, we summarize the theory used to solve for the binodals and stationary density profile of macroscopically coexisting phases of chiral active Brownian particles (cABPs). 
The theory applies the framework of Ref.~\cite{Omar2023b} which was also applied to achiral ABPs in that work. 
\subsection{\label{sec:sysdef}Microscopic Equations of Motion}
We consider $N$ particles with positional degrees of freedom $\mathbf{r}_i$ and orientational degrees of freedom $\mathbf{q}_i$. 
Then the phase space vector $\bm{\Gamma} \equiv \left[\mathbf{r}^N,\mathbf{q}^N\right]$ is a $2dN$ dimensional vector in $d$ spatial dimensions. 
The evolution of the position and orientation of the $i$th particle is given by
\begin{subequations}
\label{seq:eom}
    \begin{equation}
        \dot{\mathbf{r}}_i = U_o\mathbf{q}_i + \frac{1}{\zeta}\sum_{j\neq i}^{N}\mathbf{F}_{ij} ,\label{seq:rdot}
    \end{equation}
    \begin{equation}
        \dot{\mathbf{q}_i} = \bm{\omega}_o\times\mathbf{q}_i + \bm{\Omega}_i\times\mathbf{q}_i,\label{seq:qdot}
    \end{equation}
\end{subequations}
where $U_o$ is the active speed, $\zeta$ is the drag and $\bm{\omega}_o$ is the constant angular velocity applied to the particles.
This constant torque gives the system an overall chirality.
The noise $\bm{\Omega}_i$ has zero average and variance $2D_R\delta(t-t')\delta_{ij}\mathbf{I}$.
We interpret Eq.~\eqref{seq:qdot} in the Stratonovich convention, which is necessary to preserve the modulus of $\mathbf{q}_i$.
In the limit of vanishing noise variances, an isolated particle here will trace out a circle with radius inversely proportional to $|\bm{\omega}_o|$.
By dividing the active speed by $D_R$, one can define the run length $\ell_o \equiv U_o/D_R$ as the typical distance a particle travels before reorienting in the limit $\bm{\omega}_o\to 0$.
\subsection{\label{sec:chiralfokkerplanck}Chiral Coexistence Hierarchy of Equations}
The Fokker-Planck equation associated with Eqs.~\eqref{seq:eom} is given by:
\begin{equation}
    \label{seq:fokkerplanck}
    \frac{\partial}{\partial t}f(\bm{\Gamma};t)= \mathcal{L}f(\bm{\Gamma};t),
\end{equation}
where $f(\bm{\Gamma};t)$ is the microstate probability density and $\mathcal{L}$ is the dynamical (Fokker-Planck) operator consistent with Eq.~\eqref{seq:eom} with:
\begin{equation}
    \mathcal{L} \equiv \sum_{i=1}^{N}\left[\frac{\partial}{\partial \mathbf{r}_i}\cdot\left(-U_o\mathbf{q}_i - \frac{1}{\zeta}\sum_{j\neq i}^{N}\mathbf{F}_{ij}\right) + \boldsymbol{\nabla}_i^{R}\cdot\left( D_R\boldsymbol{\nabla}_i^R -\bm{\omega}_o \right)\right],
\end{equation}
where we have defined the rotational gradient operator as $\boldsymbol{\nabla}_i^{R} \equiv \mathbf{q}_i\times \partial /\partial \mathbf{q}_i$.
The adjoint of the dynamical operator can be similarly solved for as:
\begin{equation}
    \mathcal{L}^* \equiv \sum_{i=1}^{N}\left[\left(U_o\mathbf{q}_i + \frac{1}{\zeta}\sum_{j\neq i}^{N}\mathbf{F}_{ij}  \right)\cdot\frac{\partial}{\partial \mathbf{r}_i} + \left( D_R\boldsymbol{\nabla}_i^R +\bm{\omega}_o \right)\cdot\boldsymbol{\nabla}_i^{R}\right]\label{Ladjoint}.
\end{equation}
The average of an arbitrary macroscopic observable $\mathcal{O}$ is:
\begin{equation}
    \langle \mathcal{O} \rangle = \int\text{d}\bm{\Gamma}\mathcal{O}f(\bm{\Gamma};t) ,
\end{equation}
while the average evolution $\mathcal{O}$ can be expressed as:
\begin{equation}
    \frac{\partial \langle \mathcal{O} \rangle}{\partial t} = \int \text{d}\bm{\Gamma} \mathcal{O} \mathcal{L} f(\bm{\Gamma};t),
\end{equation}
or equivalently as:
\begin{equation}
    \frac{\partial \langle \mathcal{O} \rangle}{\partial t} = \int \text{d}\bm{\Gamma} f(\bm{\Gamma};t)\mathcal{L}^*\mathcal{O}.
\end{equation}
We now define the density $\rho$ as $\rho(\mathbf{r}) \equiv \langle \sum_{i=1}^{N}\delta(\mathbf{r}_i-\mathbf{r})\rangle$.
Then, proceeding in the same manner as Ref.~\cite{Omar2023b}, the evolution of the density is found to be:
\begin{equation}
\label{seq:denseevolve}
    \frac{\partial \rho}{\partial t} = -\boldsymbol{\nabla}\cdot\left(U_o \mathbf{m}  + \frac{1}{\zeta}\boldsymbol{\nabla}\cdot\bm{\sigma}^C\right),
\end{equation}
where $\boldsymbol{\nabla} \equiv \partial/\partial \mathbf{r}$ and we have defined the polar order density $\mathbf{m} \equiv \left\langle \sum_{i=1}^{N}\mathbf{q}_i\delta(\mathbf{r}_i-\mathbf{r})\right\rangle$ and interaction stress $\bm{\sigma}^C$ (see Refs.~\cite{Omar2023b, Langford2023} for microscopic expression).
The evolution of the density is dependent on the polar order, and an expression for the evolution of the polar order can be found in a similar fashion:
\begin{subequations}
\label{seq:polarevolve}
    \begin{equation}
        \frac{\partial\mathbf{m}}{\partial t} = -\boldsymbol{\nabla}\cdot\mathbf{J}^{m} + (1-d)D_R\mathbf{m}  + \bm{\omega}_o\times\mathbf{m},
    \end{equation}
where we have defined a flux of polar order $\mathbf{J}^{\mathbf{m}}$ as:
    \begin{equation}
        \mathbf{J}^{m} = U_o\left(\mathbf{Q} + \frac{\rho}{d}\mathbf{I}\right) + \frac{1}{\zeta}\bm{\kappa^m} + \frac{1}{\zeta}\boldsymbol{\nabla}\cdot\bm{\Sigma^m}.
    \end{equation}
\end{subequations}
Here $\bm{\kappa^m}$, $\bm{\Sigma^m}$, are the ``stress'' and ``body force''-like contributions to the flux of polar order (with microscopic expressions provided in Ref.~\cite{Omar2023b}) and we have defined the traceless nematic order tensor as $\mathbf{Q} \equiv \left\langle \sum_{i=1}^{N} \left(\mathbf{q}_i\mathbf{q}_i - \frac{1}{d}\mathbf{I}\right)\delta(\mathbf{r}_i-\mathbf{r}) \right\rangle$.
The evolution of the polar order is dependent on the traceless nematic order, and an expression for the evolution of the traceless nematic order can be found in a similar fashion:
\begin{subequations}
\label{seq:tracelessnematicevolve}
    \begin{equation}
        \frac{\partial\mathbf{Q}}{\partial t} = -\boldsymbol{\nabla}\cdot\mathbf{J}^Q  + \bm{\omega}_o\times\left(\mathbf{Q} + \frac{\rho}{d}\mathbf{I}\right) + \left[\bm{\omega}_o\times\left(\mathbf{Q} + \frac{\rho}{d}\mathbf{I}\right)\right]^T - 2dD_R\mathbf{Q},
    \end{equation}
where we have defined a flux of nematic order $\mathbf{J}^{Q}$ as:
    \begin{equation}
        \mathbf{J}^{Q} = U_o\tilde{\mathbf{B}} + \frac{1}{\zeta}\bm{\kappa^Q} + \frac{1}{\zeta}\boldsymbol{\nabla}\cdot\bm{\Sigma^Q} - \frac{U_o}{d}\mathbf{m}\mathbf{I} - \frac{1}{d\zeta} \left(\boldsymbol{\nabla}\cdot\bm{\sigma}^C\right)\mathbf{I}.
    \end{equation}
\end{subequations}
Here $\bm{\kappa^Q}$, $\bm{\Sigma^Q}$, are the ``stress'' and ``body force''-like contributions to the flux of nematic order (with microscopic expressions provided in Ref.~\cite{Omar2023b}) and we have defined $\tilde{\mathbf{B}} \equiv \left\langle \sum_{i=1}^{N} \mathbf{q}_i\mathbf{q}_i\mathbf{q}_i \delta(\mathbf{r}_i-\mathbf{r}) \right\rangle$.
Although the cross product of a second rank tensor is not defined, we have employed the following notation to describe the object which (using indicial notation) is expressed as:
\begin{equation}
    \bm{\omega}_o\times\left(\mathbf{Q} + \frac{\rho}{d}\mathbf{I}\right) + \left[\bm{\omega}_o\times\left(\mathbf{Q} + \frac{\rho}{d}\mathbf{I}\right)\right]^T = \left\langle \omega_o^{\gamma}\epsilon^{\gamma\omega\mu}\sum_{i=1}^{N}\left(\delta^{\mu\alpha}q_i^{\beta}q_i^{\omega} + \delta^{\mu\beta}q_i^{\alpha}q_i^{\omega}\right) \right \rangle.\label{seq:mtimesq}
\end{equation}
In the above indicial notation expression we use Latin subscripts to denote particle label and Greek superscripts to denote spatial component.
We find that the evolution of the traceless nematic order is dependent on $\tilde{\mathbf{B}}$. 
The coexistence theory proposed in Ref.~\cite{Omar2023b} requires a second order expansion of the dynamic stress tensor, or equivalently a third order expansion of the density flux.
Because the dynamics of the $(n)$th orientational moment is dependent on the divergence of the $(n+1)$th orientational moment, each additional level included in the hierarchy of equation serves to increase the spatial order.
Truncation of the hierarchy of equations at the nematic order is then sufficient for understanding coexistence.
\subsection{\label{sec:chiralclosure}Approximations and Closures}
The closure of the above hierarchy of equations can be greatly simplified by focusing on a system confined to two dimensions $(d=2)$.
In this case, the only relevant component of $\bm{\omega}_o$ is that in the out-of-plane direction, which we pick to be parallel to the unit vector $\mathbf{e}_z$. 
All orientations are constrained to the $(x,y)$ plane. 
We can then write $\bm{\omega}_o = \omega_o\mathbf{e}_z$.
Then two relevant inverse time scales become apparent from Eq.~\eqref{seq:eom}: $D_R$ and $\omega_o$.
We then define $\chi \equiv \omega_o/D_R$ as a dimensionless parameter capturing the relative strength of $\omega_o$ and $D_R$, which serves as a measure of the extent to which spatial parity between $x$ and $y$ has been broken. 
In two dimensions, the term $\bm{\omega}_o\times\mathbf{m}$ can be rewritten as:
\begin{equation}
    \bm{\omega}_o\times\mathbf{m} = \omega_o m_x \mathbf{e}_y - \omega_o m_y \mathbf{e}_x,
\end{equation}
and Eq.~\eqref{seq:mtimesq} can be rewritten as:
\begin{equation}
    \bm{\omega}_o\times\left(\mathbf{Q} + \frac{\rho}{d}\mathbf{I}\right) + \left[\bm{\omega}_o\times\left(\mathbf{Q} + \frac{\rho}{d}\mathbf{I}\right)\right]^T = \omega_o\begin{bmatrix}2Q_{xy} & Q_{xx} - Q_{yy} \\ Q_{xx} - Q_{yy} & 2Q_{xy}\end{bmatrix}.
\end{equation}
We now consider the steady-state. 
At steady-state, the time derivative of $\rho$, $\mathbf{m}$, and $\mathbf{Q}$ must all be zero.
Then the density-flux [the term in the parentheses of Eq.~\eqref{seq:denseevolve}] must have zero divergence.
Therefore the density-flux must be given by the curl of some vector field $\mathbf{A}$. 
We can then express the polar order as:
\begin{equation}
    \label{seq:polarordercurlA}
    \mathbf{m} = \frac{1}{U_o}\left( \boldsymbol{\nabla}\times \mathbf{A} - \frac{1}{\zeta}\boldsymbol{\nabla}\cdot\bm{\sigma}^C\right).
\end{equation}
We now close the hierarchy of equations by assuming the $\tilde{\mathbf{B}}$ field to be isotropic:
\begin{equation}
\label{seq:Bclose}
    \tilde{\mathbf{B}} = \frac{1}{4}\bm{\alpha}\cdot\mathbf{m},
\end{equation}
where $\bm{\alpha}$ is the fourth order identity tensor represented in indicial notation by $\alpha^{\mu\gamma\nu\omega} = \delta^{\mu\gamma}\delta^{\nu\omega} + \delta^{\mu\nu}\delta^{\gamma\omega} + \delta^{\mu\omega}\delta^{\nu\gamma}$.
Following Ref.~\cite{Omar2023b}, we also discard  $\bm{\Sigma^m}$, and $\bm{\Sigma^Q}$, an approximation that was found to be accurate in achiral ABPs, and introduce the following constitutive relations for $\bm{\kappa^m}$, $\bm{\kappa^Q}$:
\begin{subequations}
\label{seq:kappaapproxes}
    \begin{equation}
        \bm{\kappa^m} = -\zeta U_o\left(1 - \overline{U}(\rho)\right)\left(\mathbf{Q} + \frac{\rho}{d}\mathbf{I}\right),
    \end{equation}
    \begin{equation}
        \bm{\kappa^Q} = -\zeta U_o\left(1 - \overline{U}(\rho)\right)\tilde{\mathbf{B}},
    \end{equation}
\end{subequations}
where $\overline{U}\in [0,1]$ is the effective active speed modifier and is an equation of state provided in Section~\ref{sec:eos}.
Finally, as is standard in the active particle literature~\cite{Solon2018GeneralizedEnsembles,Omar2023b,Evans2023}, we ignore gradient contributions to the interaction stress and approximate $\bm{\sigma}^C\approx -p_C(\rho)\mathbf{I}$, where $p_C(\rho)$ is an equation of state for the interaction pressure. 
By substituting Eq.~\eqref{seq:polarordercurlA}, Eqs.~\eqref{seq:kappaapproxes}, and  Eq.~\eqref{seq:Bclose} into Eqs.~\eqref{seq:tracelessnematicevolve}; setting the time derivative of $\mathbf{Q}$ to zero; and assuming that at steady states gradients in the density are orthogonal to the residual flux $\boldsymbol{\nabla}\times\mathbf{A}$, one can solve for the components of the traceless nematic order tensor as:
\begin{subequations}
    \label{seq:Qclosed}
    \begin{align}
    Q_{xx} = -\frac{3}{16\zeta D_R}\biggl(&\frac{\partial}{\partial \rho}\left[\overline{U}\frac{\partial p_C}{\partial \rho}\right]\left(\frac{\partial \rho}{\partial x}\right)^2 +  \frac{\chi}{8}\frac{\partial}{\partial \rho}\left[\overline{U}\frac{\partial p_C}{\partial \rho}\right]\left[\left(\frac{\partial \rho}{\partial x}\right)^2 - \left(\frac{\partial \rho}{\partial y}\right)^2\right]\nonumber \\ &+\frac{\chi}{2}\frac{\partial}{\partial \rho}\left[\overline{U}\frac{\partial p_C}{\partial \rho}\right]\left(\frac{\partial \rho}{\partial x}\frac{\partial \rho}{\partial y}\right) + \overline{U}\frac{\partial p_C}{\partial \rho} \boldsymbol{\nabla}^2\rho\biggr),
    \end{align}
    \begin{align}
    Q_{yy} = -\frac{3}{16\zeta D_R}\biggl(&\frac{\partial}{\partial \rho}\left[\overline{U}\frac{\partial p_C}{\partial \rho}\right]\left(\frac{\partial \rho}{\partial y}\right)^2 +  \frac{\chi^2}{8}\frac{\partial}{\partial \rho}\left[\overline{U}\frac{\partial p_C}{\partial \rho}\right]\left[\left(\frac{\partial \rho}{\partial x}\right)^2 - \left(\frac{\partial \rho}{\partial y}\right)^2\right]\nonumber \\ &+\frac{\chi}{2}\frac{\partial}{\partial \rho}\left[\overline{U}\frac{\partial p_C}{\partial \rho}\right]\left(\frac{\partial \rho}{\partial x}\frac{\partial \rho}{\partial y}\right) + \overline{U}\frac{\partial p_C}{\partial \rho} \boldsymbol{\nabla}^2\rho\biggr),
    \end{align}
    \begin{equation}
        Q_{xy} = -\frac{3}{16\zeta D_R}\biggl(\frac{\partial}{\partial \rho}\left[\overline{U}\frac{\partial p_C}{\partial \rho}\right]\left(\frac{\partial \rho}{\partial x}\frac{\partial \rho}{\partial y}\right)  + \frac{\chi}{4}\frac{\partial}{\partial \rho}\left[\overline{U}\frac{\partial p_C}{\partial \rho}\right]\left(\left(\frac{\partial\rho}{\partial x}\right)^2 - \left(\frac{\partial\rho}{\partial y}\right)^2\right)\biggr).
    \end{equation}
\end{subequations}
We can now consider the evolution of the polar order [Eq.~\eqref{seq:polarevolve}] at steady-state:
\begin{equation}
\label{seq:polarordersub}
    \mathbf{m} = -\boldsymbol{\nabla}\cdot\left(\ell_o\overline{U}\mathbf{Q} + \frac{\ell_o\overline{U}\rho}{2}\mathbf{I}\right) + \frac{1}{D_R}\bm{\omega}_o\times\mathbf{m}.
\end{equation}
Breaking up Eq.~\eqref{seq:polarordersub} into its $x$ and $y$ components results in two equations to solve for two unknowns $m_x$ and $m_y$. 
Solving for $m_x$ and $m_y$ then recombining into vector notation results in:
\begin{equation}
    \label{seq:polarordersolve}
    \mathbf{m}\left(1 + \chi^2\right) = -\boldsymbol{\nabla}\cdot\left(\ell_o\overline{U}\mathbf{Q} + \frac{\ell_o\overline{U}\rho}{2}\mathbf{I}\right) - \frac{1}{D_R}\bm{\omega}_o\times\left(\boldsymbol{\nabla}\cdot\left(\ell_o\overline{U}\mathbf{Q} + \frac{\ell_o\overline{U}\rho}{2}\mathbf{I}\right)\right).
\end{equation}
The density flux is dependent on the divergence of the interaction stress and the polar order. 
We have approximated the interaction stress as proportional to the interaction pressure, which is given by an equation of state solely dependent on density.
Additionally, through Eqs.~\eqref{seq:Qclosed}, we have closed the traceles nematic order fully in terms of the density and equations of state.
We now have an expressions for the polar order dependent only on density, the traceless nematic order, and equations of state.
Thus we have found an expression for the density flux that is fully closed in terms of the density field:
\begin{subequations}
    \label{seq:Jclose}
    \begin{equation}
        \mathbf{J} = \frac{1}{\zeta}\boldsymbol{\nabla}\cdot\left[\mathbf{T} + \chi\mathbf{T}^{\rm cross} + \mathbf{T}^{\rm odd}\right]
    \end{equation}
    \begin{equation}
        \mathbf{T} = \begin{bmatrix}-\mathcal{P} - \frac{\zeta U_o\ell_o\overline{U} Q_{xx}}{1 + \chi^2} & -\frac{\zeta U_o\ell_o\overline{U}Q_{xy}}{1 + \chi^2} \\ -\frac{\zeta U_o\ell_o\overline{U}Q_{xy}}{1 + \chi^2} & -\mathcal{P} - \frac{\zeta U_o\ell_o\overline{U} Q_{yy}}{1 + \chi^2}  \end{bmatrix}
    \end{equation}
    \begin{equation}
        \mathbf{T}^{\rm cross} = \begin{bmatrix} \frac{\chi\zeta U_o \ell_o\overline{U}Q_{xy}}{1 + \chi^2} & -\frac{\chi\zeta U_o \ell_o\overline{U}Q_{xx}}{1 + \chi^2} \\ -\frac{\chi\zeta U_o \ell_o\overline{U}Q_{xx}}{1 + \chi^2} & -\frac{\chi\zeta U_o \ell_o\overline{U}Q_{xy}}{1 + \chi^2}\end{bmatrix}
    \end{equation}
    \begin{equation}
        \mathbf{T}^{\rm odd} = \begin{bmatrix} 0 & -\frac{\chi\zeta U_o\ell_o\overline{U}\rho}{2(1+\chi^2)} \\ \frac{\chi\zeta U_o\ell_o\overline{U}\rho}{2(1+\chi^2)} & 0\end{bmatrix}
    \end{equation}
    \begin{equation}
        \mathcal{P} = p_C(\rho) + \frac{\zeta U_o \ell_o \overline{U}\rho}{2(1 + \chi^2)}
    \end{equation}
\end{subequations}
\subsection{\label{sec:coexistencecriteria}Coexistence Criteria}
We seek the criteria at which the system macroscopically phase separates with a planar interface. 
In this limit we may restrict density gradients to be in a single direction.
Picking this direction to be $\mathbf{e}_x$, we demand that the flux in the $x$ direction is zero, i.e. $\mathbf{J}\cdot\mathbf{e}_x = 0$.
This constraint results in:
\begin{equation}
\label{seq:xfluxzero}
     0 = -\frac{\partial \mathcal{P}}{\partial \rho}\frac{\partial \rho}{\partial x} - \frac{\ell_o \zeta U_o}{(1+\chi^2)}\frac{\partial}{\partial x}\left[\overline{U}Q_{xx}\right] + \frac{\chi U_o \zeta \ell_o}{1 + \chi^2}\frac{\partial}{\partial x}\left[\overline{U}Q_{xy}\right].
\end{equation}
Integration of Eq.~\eqref{seq:xfluxzero} from $x=-\infty$ to $x=+\infty$, and noting that the bulk phases have no nematic order, results in:
\begin{equation}
\label{seq:coexist1}
    \mathcal{P}(\rho^{\rm gas}) = \mathcal{P}(\rho^{\rm liq}).
\end{equation}
\textit{Indefinite} integration of Eq.~\eqref{seq:xfluxzero} results in:
\begin{equation}
\label{seq:xfluxzeroindefinteg}
    \mathcal{P}(\rho) - \mathcal{P}^{\rm coexist} = -\frac{\ell_o\zeta U_o\overline{U}Q_{xx}}{1 + \chi^2} + \frac{\chi\ell_o\zeta U_o\overline{U}Q_{xy}}{1 + \chi^2},
\end{equation}
where we have identified the constant of integration as the coexistence pressure $\mathcal{P}^{\rm coexist}$.
Substitution of Eqs.~\eqref{seq:Qclosed} into Eq.~\eqref{seq:xfluxzeroindefinteg} results in:
\begin{align}
\label{seq:xfluxzeroindefintegpredef}
    \mathcal{P}(\rho) - \mathcal{P}^{\rm coexist}  =  &\frac{3\ell_o^2}{16(1+\chi^2)}\left(\overline{U}\right)^2\frac{\partial p_C}{\partial \rho}\frac{\partial^2 \rho}{\partial x^2} \nonumber \\ & +  \left(1 - \frac{\chi^2}{8}\right)\frac{3\ell_o^2}{16(1+\chi^2)}\overline{U}\frac{\partial }{\partial \rho}\left[\overline{U}\frac{\partial p_C}{\partial \rho}\right]\left(\frac{\partial \rho}{\partial x}\right)^2. 
\end{align}
We now define the following constants:
\begin{subequations}
    \label{seq:coexistinterfacialcoeff}
    \begin{equation}
        \mathfrak{a}(\rho) = \frac{3\ell_o^2}{16(1+\chi^2)}\left(\overline{U}\right)^2\frac{\partial p_C}{\partial \rho},
    \end{equation}
    \begin{equation}
        \mathfrak{b}(\rho) = \left(1 - \frac{\chi^2}{8}\right)\frac{3\ell_o^2}{16(1+\chi^2)}\overline{U}\frac{\partial}{\partial \rho}\left[\overline{U}\frac{\partial p_C}{\partial \rho}\right],
    \end{equation}
\end{subequations}
where substitution of these constants into Eq.~\eqref{seq:xfluxzeroindefintegpredef} results in the compact notation:
\begin{equation}
\label{seq:xfluxintegrateniceform}
    \mathcal{P}(\rho) - \mathcal{P}^{\rm coexist}  = \mathfrak{a}(\rho)\frac{\partial^2\rho}{\partial x^2} + \mathfrak{b}(\rho)\left(\frac{\partial \rho}{\partial x}\right)^2.
\end{equation}
Eq.~\eqref{seq:xfluxintegrateniceform} is in a form amenable to the macroscopic coexistence theory of Ref.~\cite{Omar2023b}. 
We may define the pseudovariable:
\begin{equation}
\label{seq:pseudodef}
    \frac{\partial^2 \mathcal{E}}{\partial \rho^2} = \frac{2\mathfrak{b}(\rho) - \frac{\partial\mathfrak{a}}{\partial \rho}}{\mathfrak{a}(\rho)}\frac{\partial\mathcal{E}}{\partial \rho}.
\end{equation}
Substitution of Eqs.~\eqref{seq:coexistinterfacialcoeff} into Eq.~\eqref{seq:pseudodef} and solution of the differential equation results in:
\begin{equation}
    \frac{\partial\mathcal{E}}{\partial \rho}\sim \left(\overline{U}\right)^{-\frac{\chi^2}{4}}\left(\frac{\partial p_C}{\partial \rho}\right)^{1 - \frac{\chi^2}{4}}.
\end{equation}
In the limit $\chi\to 0$, the pseudovariable recovers the same solutions as found for achiral ABPs~\cite{Omar2023b,Evans2023,Langford2023}.
Integration of Eq.~\eqref{seq:xfluxintegrateniceform} with respect to the pseudovariable from one binodal to the other results in:
\begin{equation}
\label{seq:coexist2}
    \int_{\rho^{\rm gas}}^{\rho^{\rm liq}}\left[\mathcal{P}(\rho) - \mathcal{P}^{\rm coexist}\right]\frac{\partial \mathcal{E}}{\partial \rho}\text{d}\rho.
\end{equation}
Together, Eqs.~\eqref{seq:coexist1} and~\eqref{seq:coexist2} are the coexistence criteria for chiral ABPs. 
They represent two equations for which two unknowns, the coexisting densities in each phase, can be solved for. 
Further, as we discussed in Ref.~\cite{Langford2023} and extended to finite-size phases in Ref.~\cite{Langford2024TheMatter}, the coexistence criteria can be used to solve for the density profile connecting the two phases.
\subsection{\label{sec:tangflux}Steady-State Tangential Flux}
\begin{figure}
	\centering
	\includegraphics[width=0.75\textwidth]{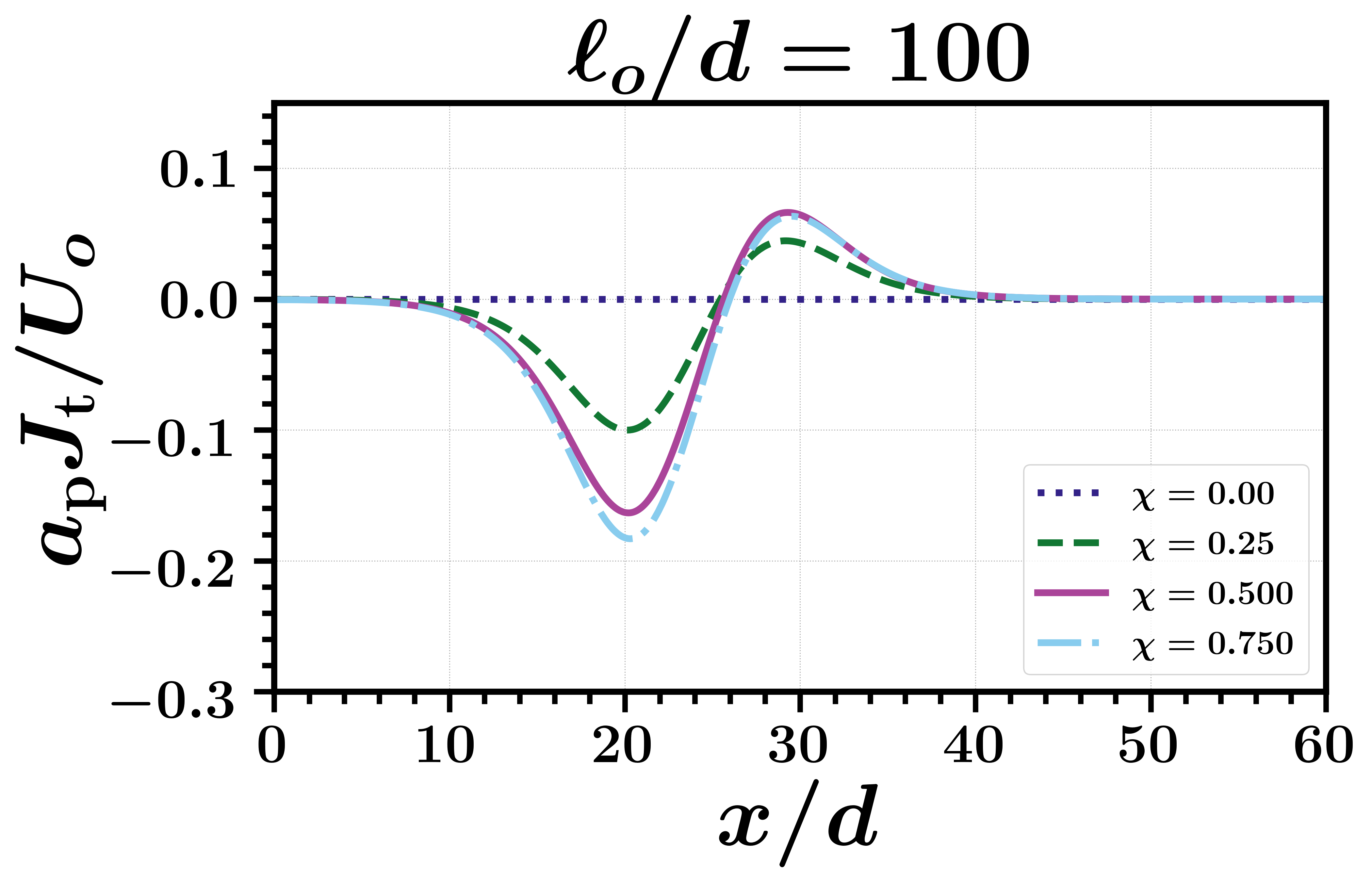}
	\caption{\protect\small{{Theoretically predicted flux tangential to the interface for $\ell_o/d = 75$ and selected values of $\chi$. Here $a_p = \pi d^2 /4$ is the area of a particle and $J_t = \mathbf{J}\cdot\mathbf{e}_y$ is the tangential flux.}}}
	\label{sfig:tanflux75}
\end{figure}

\begin{figure}
	\centering
	\includegraphics[width=0.75\textwidth]{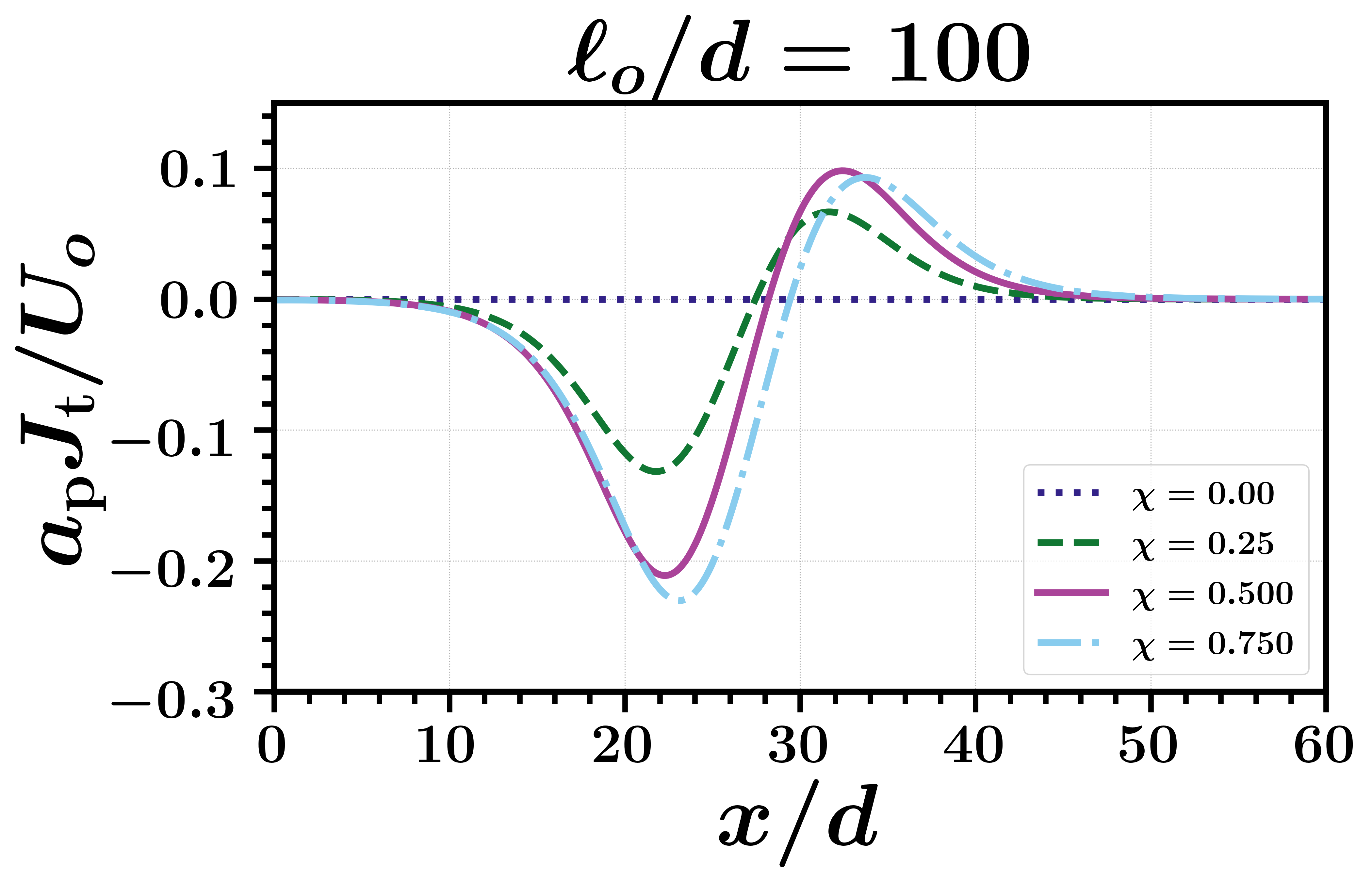}
	\caption{\protect\small{{Theoretically predicted flux tangential to the interface for $\ell_o/d = 100$ and selected values of $\chi$. Here $a_p = \pi d^2 /4$ is the area of a particle and $J_t = \mathbf{J}\cdot\mathbf{e}_y$ is the tangential flux. }}}
	\label{sfig:tanflux100}
\end{figure}

\begin{figure}
	\centering
	\includegraphics[width=0.75\textwidth]{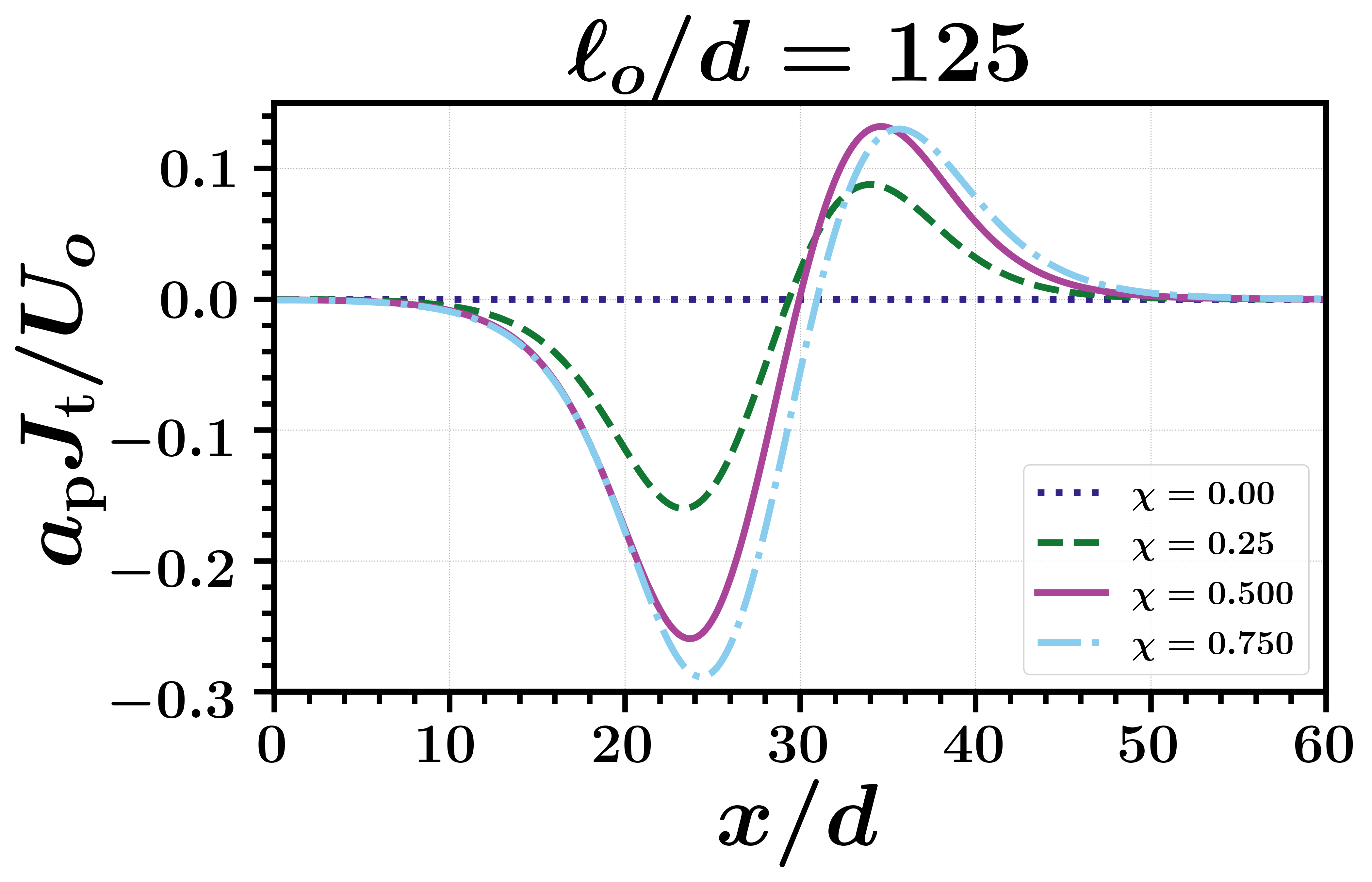}
	\caption{\protect\small{{Theoretically predicted flux tangential to the interface for $\ell_o/d = 125$ and selected values of $\chi$. Here $a_p = \pi d^2 /4$ is the area of a particle and $J_t = \mathbf{J}\cdot\mathbf{e}_y$ is the tangential flux. }}}
	\label{sfig:tanflux125}
\end{figure}

While macroscopic coexistence demands that $\mathbf{J}\cdot\mathbf{e}_x = 0$, the tangential flux $\mathbf{J}\cdot\mathbf{e}_y$ may be non-zero. 
For a system with no gradients in the $y$ direction, the tangential flux is found to be:
\begin{equation}
    \mathbf{J}\cdot\mathbf{e}_y = -\frac{ U_o \ell_o }{1 + \chi^2}\frac{\partial}{\partial x}\left[Q_{xy}\overline{U} + \chi Q_{xx}\overline{U}\right] + \frac{\chi U_o \ell_o}{2(1+\chi^2)}\frac{\partial}{\partial x}\left[\overline{U}\rho\right].
\end{equation}
Noting that $Q_{xy}=0$ when $\chi=0$, the tangential flux vanishes in the limit $\chi\to 0$. 
Therefore tangential fluxes at steady state in ABPs can only be supported upon violation of spatial parity symmetry.
Substitution of the closure relations for $\mathbf{Q}$ [Eq.~\eqref{seq:Qclosed}] and the density profiles which may be solved for using the coexistence criteria results will yield steady-state tangential flux profiles, which we report in Figs.~\ref{sfig:tanflux75}-~\ref{sfig:tanflux125}
\subsection{\label{sec:eos}Equations of State}
Evaluation of the steady-state binodals, density profiles, and tangential fluxes requires equations of state for the interaction pressure $p_C(\rho)$ and the dimensionless active speed $\overline{U}$.
For simplicity, we choose equations of state that capture the expected qualitative behavior while satisfying known physical limits. 
For example, in the limit of zero area fraction we demand that the interaction pressure is zero and the dimensionless active speed is one. 
In the limit of close-packing ($\phi \rightarrow \phi_o$) we demand that the interaction pressure diverges and the effective active speed modifier is zero.
We therefore adopt the interaction pressure proposed in Ref.~\cite{Takatori2015TowardsMatter}:
\begin{equation}
    \frac{p_C d}{\zeta U_o} = \frac{8}{\pi}\phi^2\left(1 - \frac{\phi}{\phi_o}\right)^{-1},
\end{equation}
where $\phi \equiv \pi \rho d^2 /4$ is the area fraction and a value of $0.9$ is used for the close-packed area fraction $\phi_o$.
We use the following form for the effective active speed:
\begin{equation}
    \overline{U} = \left[1 + \phi\left(1 - \frac{\phi}{\phi_o}\right)^{-1}\right]^{-1}.
\end{equation} \newpage
\section{\label{sec:capillarity}Interfacial Dynamics of Coexisting Chiral Active Particles}
In this Section, we summarize the theory used to model the interfacial dynamics of coexisting phases of chiral ABPs.
First we present a fluctuating hydrodynamic description of chiral ABPs which can be derived using the procedure reported by Ref.~\cite{Langford2023}.
We then follow the process of Refs.~\cite{Bray1994TheoryKinetics,Fausti2021CapillarySeparation,Langford2023} to connect the stochastic description of the density dynamics to a Langevin equation for the interfacial height. 
From this Langevin equation we define key quantities such as the capillary tension and odd flow coefficient.
We then perform a linear stability analysis of this Langevin equation, demonstrating that the interfacial dynamics are stable as long as the capillary tension is positive.
Finally, we analyze the stationary fluctuations and distribution of the interface, finding that only the capillary tension is relevant for the steady-state properties of the interface.
\subsection{\label{sec:flucthydro}Fluctuating Hydrodynamics}
The fluctuating hydrodynamics of achiral ABPs were derived in Ref.~\cite{Langford2023}, following the procedures set by Refs.~\cite{Dean1996LangevinProcesses,Cugliandolo2015StochasticDipoles}.
These dynamics, valid when the magnitude of the flux $\mathbf{J}$ is small and on timescales larger than $1/D_R$, can straightforwardly be extended to include the external torque $\mathbf{M}_o$.
Ref.~\cite{Langford2023} found that the deterministic terms of the fluctuating dynamics would be exactly the same as those found via the Fokker-Planck equation, except sampled by a coarse-graining kernel $\Delta(\mathbf{r}-\mathbf{r}_i)$ rather than ensemble-averaged.
Then the fluctuating hydrodynamics of chiral ABPs is given by:
\begin{subequations}
    \label{seq:flucthydro}
    \begin{equation}
        \frac{\partial \rho}{\partial t} = -\boldsymbol{\nabla}\cdot\mathbf{J},
    \end{equation}
    \begin{equation}
        \mathbf{J} = \frac{1}{\zeta}\left(\mathbf{T} + \mathbf{T}^{\rm cross} + \mathbf{T}^{\rm odd}\right) + \bm{\eta}^{\rm act},
    \end{equation}
where $\mathbf{T}$, $\mathbf{T}^{\rm cross}$, and $\mathbf{T}^{\rm odd}$ were provided in Section~\ref{sec:chiralclosure} and $\bm{\eta}^{\rm act}$ is a zero-average noise with covariance:
    \begin{equation}
        \langle \bm{\eta}^{\rm act}(\mathbf{r},t)\bm{\eta}^{\rm act}(\mathbf{r}',t') \rangle = 2\frac{k_BT^{\rm act}}{\zeta}\left(\rho\mathbf{I} - 2\mathbf{Q}\right)\delta(t-t')\delta(\mathbf{r}-\mathbf{r}').
    \end{equation}
\end{subequations}
Here $k_BT^{\rm act}\equiv \zeta U_o \ell_o/2$ is the athermal energy scale of the fluctuations.
It is worth  noting that although $\mathbf{T}^{\rm odd}$ enters the flux, it will never affect the overall density evolution as two divergences of an antisymmetric tensor must be zero.
Despite $\mathbf{T}^{\rm odd}$ not impacting the overall density evolution, the presence of parity-violating terms in $\mathbf{T}$, $\mathbf{T}^{\rm cross}$ such as $\partial_x\rho\partial_y\rho$
It is convenient to express $\mathbf{T}$ in terms of derivatives of the density field. 
This can be done by substituting the closure of the traceless nematic order Eq.~\eqref{seq:Qclosed} into Eq.~\eqref{seq:Jclose}:
\begin{align}
    \mathbf{T} = &\left[-\mathcal{P}(\rho) + a(\rho)\boldsymbol{\nabla}^2\rho + \frac{\chi}{2}b(\rho)\frac{\partial\rho}{\partial x}\frac{\partial \rho}{\partial y} +\frac{\chi^2}{8}b(\rho)\left[\left(\frac{\partial \rho}{\partial x}\right)^2-\left(\frac{\partial \rho}{\partial y}\right)^2\right]\right]\mathbf{I} \nonumber \\ & + b(\rho)\boldsymbol{\nabla}\rho\boldsymbol{\nabla}\rho + \frac{\chi}{4}b(\rho)\left[\left(\frac{\partial \rho}{\partial x}\right)^2-\left(\frac{\partial \rho}{\partial y}\right)^2\right]\left(\mathbf{e}_x\mathbf{e}_y + \mathbf{e}_y\mathbf{e}_x\right),
\end{align}
where we have defined
\begin{equation}
    a(\rho) = \frac{3\ell_o^2}{16(1+\chi^2)}\overline{U}^2\frac{\partial p_C}{\partial \rho},
\end{equation}
and
\begin{equation}
    b(\rho) = \frac{3\ell_o^2}{16(1+\chi^2)}\overline{U}\frac{\partial}{\partial \rho}\left[\overline{U}\frac{\partial p_C}{\partial \rho}\right].
\end{equation}
Additionally, $\mathbf{T}^{\rm cross}$ can be compactly expressed as:
\begin{equation}
    \mathbf{T}^{\rm cross} = T_{xy}\left(\mathbf{e}_y\mathbf{e}_y - \mathbf{e}_x\mathbf{e}_x\right) + \chi\left( T_{xx} + \mathcal{P}\right)\left(\mathbf{e}_x\mathbf{e}_y + \mathbf{e}_y\mathbf{e}_x\right)
\end{equation}
\subsection{\label{sec:interfaceeom}Interfacial Langevin Dynamics}
We connect the dynamics of the density field to the dynamics of an interfacial height field via the ansatz~\cite{Bray2002InterfaceShear}:
\begin{equation}
\label{seq:ansatz}
    \rho(\mathbf{r},t) = \upvarphi[x - h(y)],
\end{equation}
where $\upvarphi$ is the stationary density profile solved for from the coexistence criteria. 
We proceed in the same manner as Ref.~\cite{Langford2023}, substituting Eq.~\eqref{seq:ansatz} into Eqs.~\eqref{seq:flucthydro}, Fourier transforming in the $y$ direction, and applying the Green's function of the Laplace operator.
The pseudovariable which eliminates nonlinear terms in the interfacial dynamics is found to be different than the one used for the generalized Maxwell construction:
\begin{equation}
    \frac{\partial\mathcal{E}_{\rm cw}}{\partial \rho}\sim \left(\overline{U}\right)^{\frac{\chi^2}{4}} \left(\frac{\partial p_C}{\partial \rho}\right)^{1 + \frac{\chi^2}{4}}.
\end{equation}
The pseudovariable used to construct the binodals was chosen to eliminate nonlinear gradients in the $x$ direction while the pseudovariable used to derive the interfacial dynamics was chosen to eliminate nonlinear gradients in the $y$ direction. 
The terms responsible for the $\chi^2/4$ exponent in the pseudovariable came from a term which in the flux is proportional to a \textit{difference} in the squared spatial gradients.
Thus the pseudovariable must have a corresponding antisymmetry for eliminating nonlinear gradients perpendicular to the interface as opposed to tangential to the interface.
The difference in squared derivative terms leading to this antisymmetry are only possible in systems that violate spatial parity.
Following integration with respect to $\mathcal{E}_{\rm cw}$, we find the following interfacial dynamics:
\begin{subequations}
    \begin{equation}
    \label{seq:interfacelangevin}
        \zeta_{\rm eff}\frac{\partial h}{\partial t} = -k^2|k|\upgamma_{\rm cw}h - ik|k|\nu_{\rm odd}h + \xi^{\rm iso} + \xi^{\rm aniso} + \mathcal{O}\left(h^2k^3\right),
    \end{equation}
where we have defined the capillary tension $\upgamma_{\rm cw}$, the odd flow coefficient $\nu_{\rm odd}$, an effective drag $\zeta_{\rm eff}$, and the noises $\chi^{\rm iso/aniso}$ which arise from the result of the described mathematical procedure on $-\boldsymbol{\nabla}\cdot\bm{\eta}^{\rm act}$. 
The capillary tension $\upgamma_{\rm cw}$ is found to be:
    \begin{equation}
        \upgamma_{\rm cw} = \frac{A(k)}{\rho^{\rm surf}B(k)} \left[ \int dx \frac{\partial\mathcal{E}_{\rm cw}}{\partial x}a(\rho)\upvarphi' - \left(1 - \frac{\chi^2}{2}\right)\int dx \frac{\partial\mathcal{E}_{\rm cw}}{\partial x}\int dx' e^{-|k||x-x'|}\text{sgn}(x-x') b(\rho)(\upvarphi')^2\right],
    \end{equation}
the odd flow coefficient $\nu_{\rm odd}$ is found as:
    \begin{align}
        \nu_{\rm odd} = \frac{A(k)}{\rho^{\rm surf}B(k)}\Biggl[& \frac{\chi}{2}\int dx \frac{\partial\mathcal{E}_{\rm cw}}{\partial x} b(\rho)(\upvarphi')^2 +|k|\chi\int dx \frac{\partial\mathcal{E}_{\rm cw}}{\partial x}\int dx' e^{-|k||x-x'|}b(\rho)(\upvarphi')^2 \nonumber \\ &- k\chi\int dx \frac{\partial\mathcal{E}_{\rm cw}}{\partial x}\int dx' e^{-|k||x-x'|} sgn(x-x')a(\rho)\upvarphi' \Biggr],
    \end{align}
and we have defined the effective drag as:
    \begin{equation}
        \zeta_{\rm eff} = \frac{\zeta A^2(k)}{2B(k) \rho^{\rm surf}},
    \end{equation}
where $\rho^{\rm surf}\equiv (\rho^{\rm liq} + \rho^{\rm gas})/2$. 
The functions $A(k)$ and $B(k)$ are defined as:
\begin{equation}
    A(k) = \int dx' \int dx e^{-|k||x-x'|}\upvarphi'(x) \frac{\mathcal{E}_{\rm cw}}{\partial x'},
\end{equation}
\begin{equation}
    B(k) = \int dx' \int dx e^{-|k||x-x'|}\frac{\partial \mathcal{E}_{\rm cw}}{\partial x}\frac{\mathcal{E}_{\rm cw}}{\partial x'}.
\end{equation}
The noises are uncorrelated, with zero average and variances:
\begin{equation}
    \langle \xi^{\rm iso}(k,t)\xi^{\rm iso}(k',t') \rangle = 2|k|k_BT^{\rm act}\zeta_{\rm eff}(2\pi)\delta(t-t')\delta(k+k'),\label{seq:isonoisecorr}
\end{equation}
and
\begin{equation}
    \langle \xi^{\rm aniso}(k,t)\xi^{\rm aniso}(k',t') \rangle = \frac{\zeta_{\rm eff}\left(C(k) + D(k) + R(k)\right)}{\rho^{\rm surf}B(k)}(2\pi)\delta(t-t')\delta(k+k').\label{seq:anisonoisecorr}
\end{equation}
Here we have defined the functions $C(k)$, $D(k)$, and $R(k)$ as:
\begin{align}
    C(k) = \int\int dudu'&\frac{\partial\mathcal{E}_{\rm cw}}{\partial u}\frac{\partial \mathcal{E}_{\rm cw}}{\partial u'}\int\int dxdx'\nonumber \\ &  \times  e^{-|k||u-x|}e^{-|k||u'-x'|}\left(\frac{\upvarphi''k^2a(\rho)}{\overline{U}}\delta(x-x') - \frac{\partial}{\partial x}\left(\frac{\upvarphi''a(\rho)}{\overline{U}}\frac{\partial}{\partial x}\delta(x-x')\right)\right),
\end{align}
\begin{align}
    D(k) = \int\int dudu'&\frac{\partial\mathcal{E}_{\rm cw}}{\partial u}\frac{\partial \mathcal{E}_{\rm cw}}{\partial u'}\int\int dxdx' \nonumber \\ &\times  e^{-|k||u-x|}e^{-|k||u'-x'|}\frac{\partial}{\partial x}\left(\frac{(\upvarphi')^2b(\rho)}{\overline{U}}\frac{\partial}{\partial x}\delta(x-x')\right),
\end{align}
\begin{align}
    R(k) = &\int\int dudu'\frac{\partial\mathcal{E}_{\rm cw}}{\partial u}\frac{\partial \mathcal{E}_{\rm cw}}{\partial u'}\int\int dxdx' \nonumber  \\ &\times  e^{-|k||u-x|}e^{-|k||u'-x'|}\left(\frac{(\upvarphi')^2k^2\chi^2b(\rho)}{8\overline{U}}\delta(x-x') - \frac{\partial}{\partial x}\left(\frac{(\upvarphi')^2\chi^2b(\rho)}{8\overline{U}}\frac{\partial}{\partial x}\delta(x-x')\right)\right)
\end{align}
\end{subequations}
\subsubsection{\label{sec:captensionplots}Interfacial Coefficient Plots}
\begin{figure}
	\centering
	\includegraphics[width=0.75\textwidth]{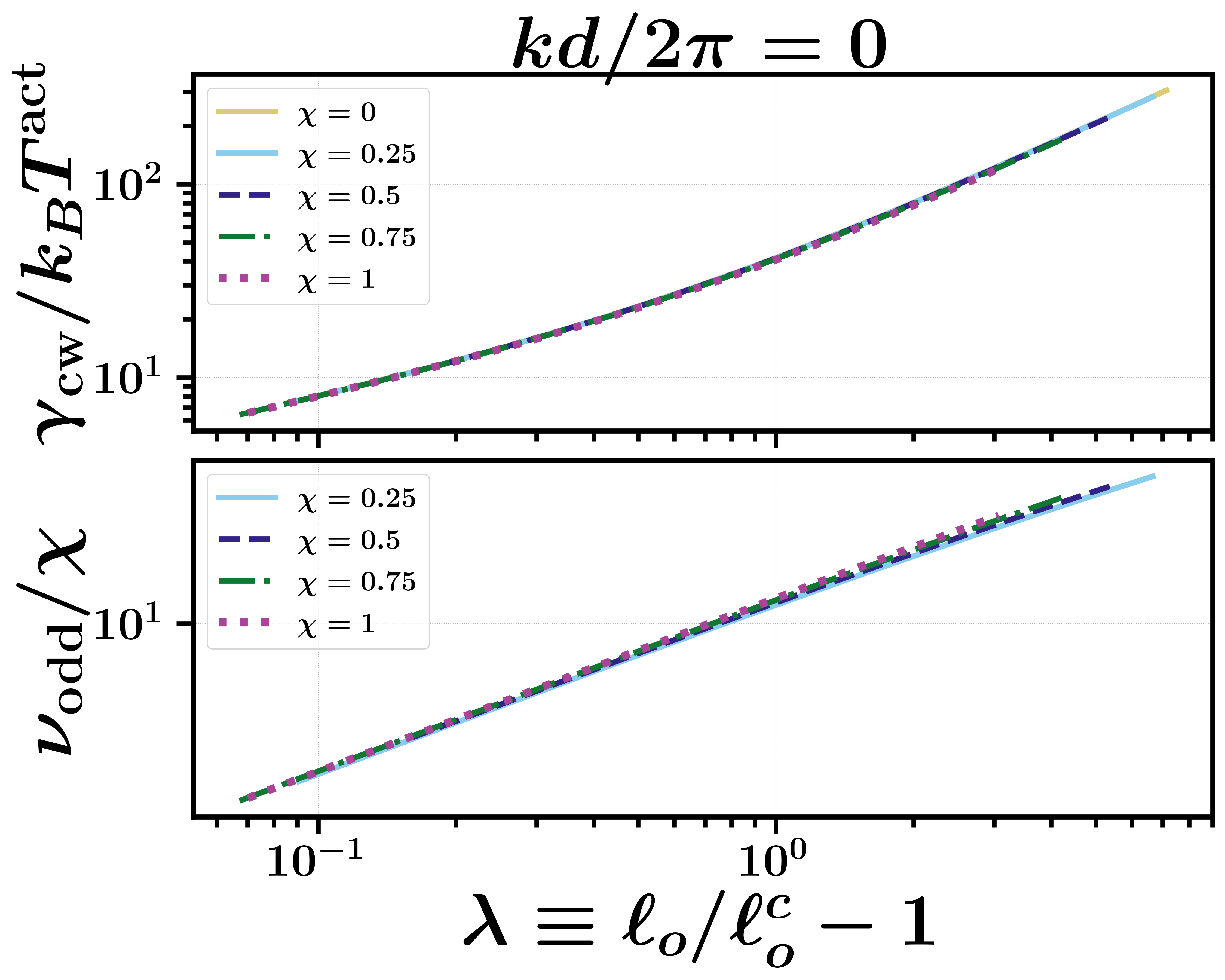}
	\caption{\protect\small{{$k\to 0$ limit of $\upgamma_{\rm cw}/k_BT^{\rm act}$ and $\nu_{\rm odd}/\chi$ as a function of reduced run length. }}}
	\label{sfig:chiraltension0k}
\end{figure}

\begin{figure}
	\centering
	\includegraphics[width=0.75\textwidth]{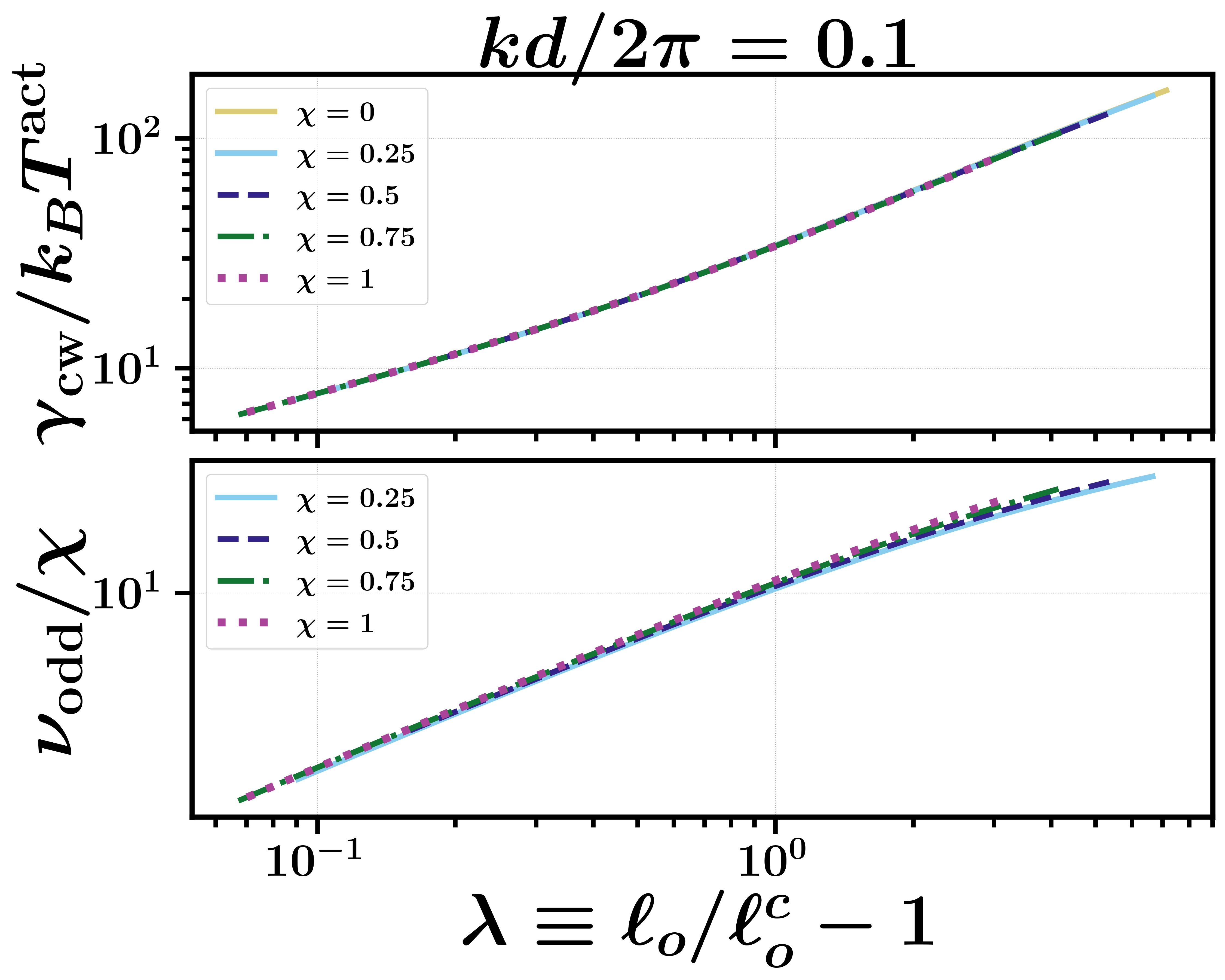}
	\caption{\protect\small{{$\upgamma_{\rm cw}/k_BT^{\rm act}$ and $\nu_{\rm odd}/\chi$  when $kd/2\pi=0.1$ as a function of reduced run length. }}}
	\label{sfig:chiraltension01k}
\end{figure}

\begin{figure}
	\centering
	\includegraphics[width=0.75\textwidth]{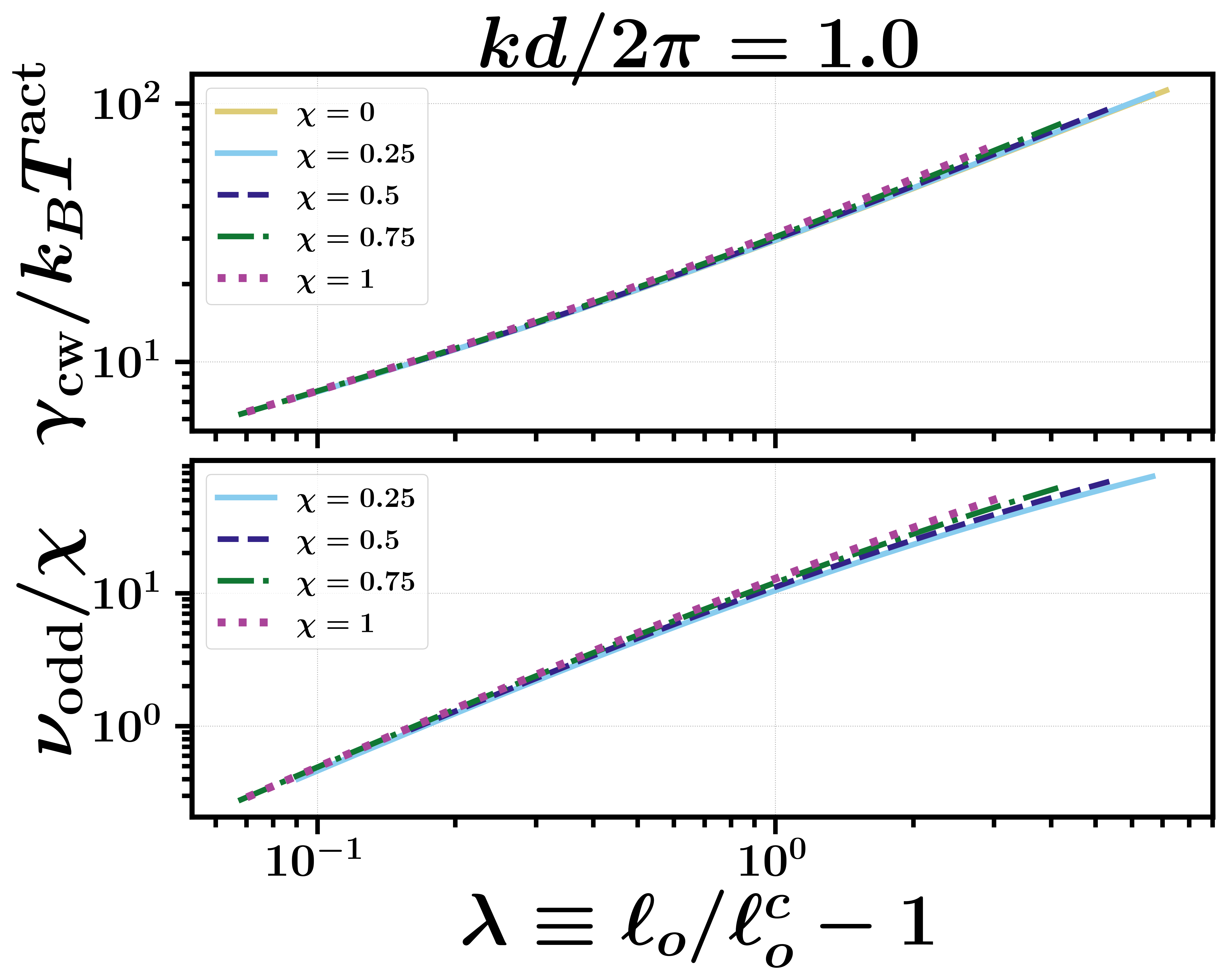}
	\caption{\protect\small{{$\upgamma_{\rm cw}/k_BT^{\rm act}$ and $\nu_{\rm odd}/\chi$  when $kd/2\pi=1.0$ as a function of reduced run length. }}}
	\label{sfig:chiraltension1k}
\end{figure}

The interfacial coefficients $\upgamma_{\rm cw}$ and $\nu_{\rm odd}$ are functions of $\ell_o$, $\chi$, and $k$.
As discussed in the main text, we expect $\upgamma_{\rm cw}/k_BT^{\rm act}$ and $\nu_{\rm odd}$ to collapse across all values of $\chi$ when plotted against reduced run length $\lambda = \ell_o/\ell_o^c - 1$.
We show these plots for $kd/2\pi=\{0,0.1,1.0\}$ in Fig.~\ref{sfig:chiraltension0k}, Fig.~\ref{sfig:chiraltension01k}, and Fig.~\ref{sfig:chiraltension1k}.
The predicted collapse is most robust for $k \rightarrow 0$ but continues to be approximately correct even at the highest $k$s.
\subsection{\label{sec:linstability}Linear Stability Analysis}
We now seek to determine whether the \textit{deterministic} terms of Eq.~\eqref{seq:interfacelangevin} are stable against the fluctuations induced by the stochastic terms. 
Breaking Eq.~\eqref{seq:interfacelangevin} into real and imaginary components, we find:
\begin{subequations}
    \begin{equation}
        \frac{\partial \text{Re}\left(h\right)}{\partial t} = -k^2|k|\upgamma_{\rm cw}\text{Re}\left(h\right) + k|k|\nu_{\rm odd}\text{Im}\left(h\right),
    \end{equation}
    \begin{equation}
        \frac{\partial \text{Im}\left(h\right)}{\partial t} =  - k|k|\nu_{\rm odd}\text{Re}\left(h\right) -k^2|k|\upgamma_{\rm cw}\text{Im}\left(h\right).
    \end{equation}
\end{subequations}
The fixed point of these equations is $\text{Re}\left(h\right) = \text{Im}\left(h\right) = 0$.
Then, rewriting $h= 0 + \delta h$, we find the dynamics of a small perturbation to the fixed point as:
\begin{subequations}
    \begin{equation}
        \frac{\partial \text{Re}\left(\delta h\right)}{\partial t} = -k^2|k|\upgamma_{\rm cw}\text{Re}\left(\delta h\right) + k|k|\nu_{\rm odd}\text{Im}\left(\delta h\right),
    \end{equation}
    \begin{equation}
        \frac{\partial \text{Im}\left(\delta h\right)}{\partial t} =  - k|k|\nu_{\rm odd}\text{Re}\left(\delta h\right) -k^2|k|\upgamma_{\rm cw}\text{Im}\left(\delta h\right).
    \end{equation}
\end{subequations}
The stability of the fixed point $h=0$ can then be found by solving for the eigenvalues of the following matrix:
\begin{equation}
    \begin{bmatrix} -k^2|k|\upgamma_{\rm cw} &  k|k|\nu_{\rm odd} \\ - k|k|\nu_{\rm odd} & -k^2|k|\upgamma_{\rm cw}\end{bmatrix}.
\end{equation}
This matrix has two complex eigenvalues: $-k^2|k|\upgamma_{\rm cw} \pm ik|k|\nu_{\rm odd}$.
The real parts of these eigenvalues are negative on the condition $\upgamma_{\rm cw}$ is positive.
Therefore Eq.~\eqref{seq:interfacelangevin} is linearly stable against fluctuations provided $\upgamma_{\rm cw} > 0$.
\subsection{\label{sec:fluctspectra}Fluctuation Spectra}
We now seek to describe the correlations  of $h(k)$. 
We begin with Eq.~\eqref{seq:interfacelangevin} evaluated at $+k$ and multiply by $h(-k)$:
\begin{align}
\label{seq:hkhnegk}
    \zeta_{\rm eff}\frac{\partial h(k)}{\partial t}h(-k) = -k^2|k|\upgamma_{\rm cw}h(k)h(-k) - ik|k|\nu_{\rm odd}h(k)h(-k) + (\xi^{\rm iso}(k) + \xi^{\rm aniso}(k))h(-k),
\end{align}
where we consider are considering low $k$ and thus the $k$ dependence of $\upgamma_{\rm cw}$, $\zeta_{\rm eff}$, and $\nu_{\rm odd}$ are negligible. 
We also consider Eq.~\eqref{seq:interfacelangevin} evaluated at $-k$ and multiply by $h(+k)$:
\begin{align}
\label{seq:hnegkhk}
    \zeta_{\rm eff}\frac{\partial h(-k)}{\partial t}h(k) = -k^2|k|\upgamma_{\rm cw}h(-k)h(k) + ik|k|\nu_{\rm odd}h(-k)h(k) + (\xi^{\rm iso}(-k) + \xi^{\rm aniso}(-k))h(k).
\end{align}
Adding Eq.~\eqref{seq:hkhnegk} to Eq.~\eqref{seq:hnegkhk} results in:
\begin{equation}
    \zeta_{\rm eff}\frac{\partial}{\partial t}\left(h(k)h(-k)\right) = -2k^2|k|\upgamma_{\rm cw}h(k)h(-k) + (\xi^{\rm iso}(-k) + \xi^{\rm aniso}(-k))h(k)+(\xi^{\rm iso}(k) + \xi^{\rm aniso}(k))h(-k).
\end{equation}
Averaging over the noise and taking the steady state results in:
\begin{equation}
    2k^2|k|\upgamma_{\rm cw}\langle |h(k)|^2\rangle = \langle (\xi^{\rm iso}(-k) + \xi^{\rm aniso}(-k))h(k)\rangle+\langle(\xi^{\rm iso}(k) + \xi^{\rm aniso}(k))h(-k)\rangle.
\end{equation}
The implicit solution to Eq.~\eqref{seq:interfacelangevin} is given by:
\begin{align}
    h(k,t) = &h(k,0)\text{exp}\left[\frac{-t\left(k^2|k|\upgamma_{\rm cw} + ik|k|\nu_{\rm odd}\right)}{\zeta_{\rm eff}}\right] \nonumber \\ &+ \frac{1}{\zeta}\int_0^{t'}dt' \exp\left[\frac{-(t-t')(k^2|k|\upgamma_{\rm cw} + ik|k|\nu_{\rm odd})}{\zeta_{\rm eff}}\right]\left(\xi^{\rm iso}(k,t')+\xi^{\rm aniso}(k,t')\right)
\end{align}
We now multiply by $(\xi^{\rm iso}(-k,t) + \xi^{\rm aniso}(-k,t))$ and average, finding:
\begin{equation}
    \langle h(k,t) (\xi^{\rm iso}(-k,t) + \xi^{\rm aniso}(-k,t)) \rangle = |k|k_BT^{\rm act} L + \frac{1}{2}\mathcal{A}(k),
\end{equation}
where we have defined:
\begin{equation}
    \mathcal{A}(k) = L\frac{C(k) + D(k) +R(k)}{\rho^{\rm surf}B(k)}.
\end{equation}
Then it is straightforward to show that:
\begin{equation}
    \langle |h(k)|^2 \rangle = \frac{L|k|k_BT^{\rm act} + 0.5\mathcal{A}(k)}{k^2|k|\upgamma_{\rm cw}},
\end{equation}
and in the limit $k\ell_o << 1$, where $\mathcal{A}$ was found to be negligible~\cite{Langford2023}:
\begin{equation}
    \langle |h(k)|^2 \rangle = \frac{Lk_BT^{\rm act}}{k^2\upgamma_{\rm cw}}.\label{seq:fluctspectrum}
\end{equation}
\subsubsection{\label{sec:particlesimspec}Simulation Spectra}
\begin{figure}
	\centering
	\includegraphics[width=0.75\textwidth]{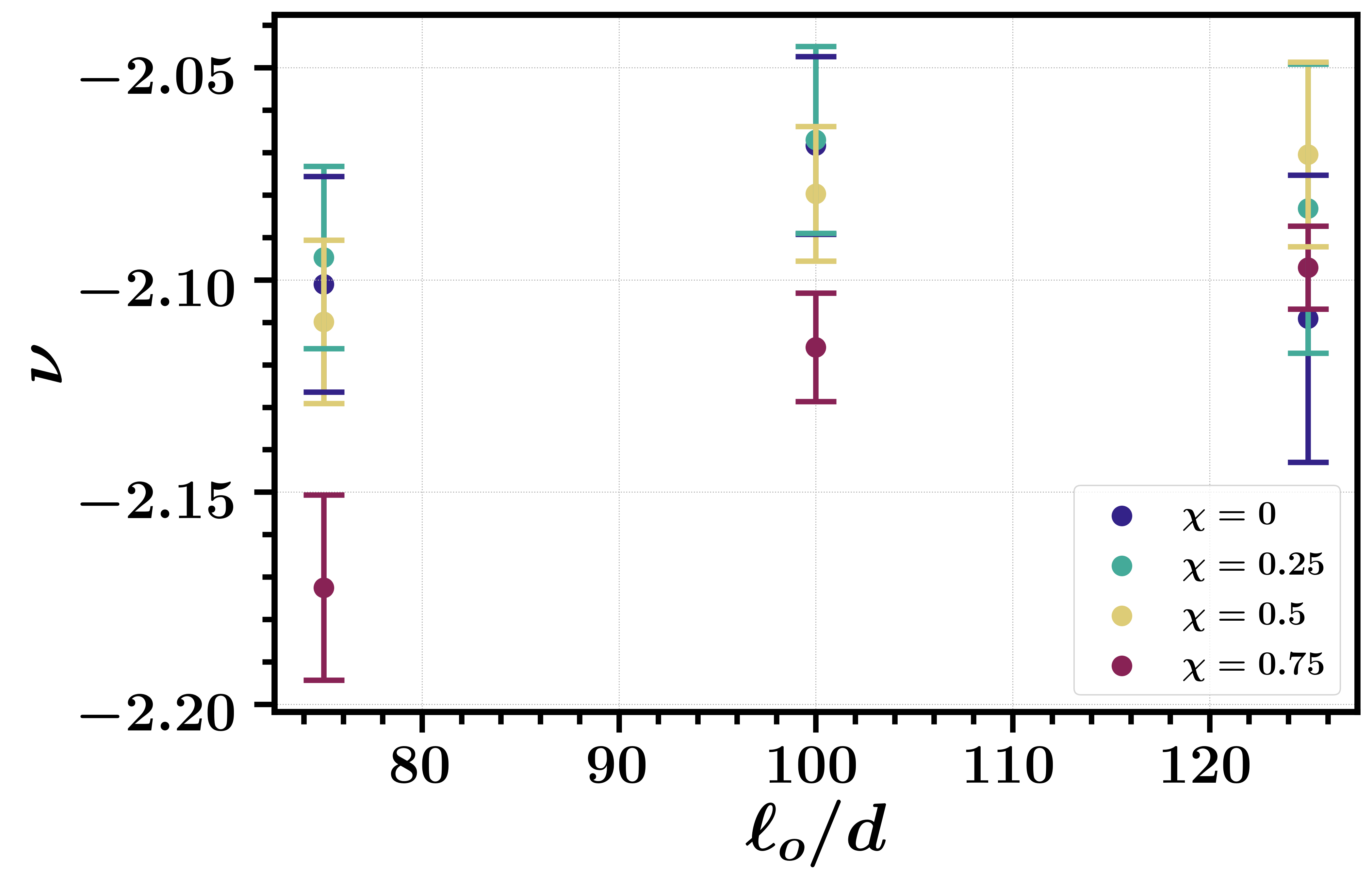}
	\caption{\protect\small{{$\nu$ as measured from a fit of the fluctuation spectrum to $\langle |h(k)|^2\rangle = K_s k^{\nu}$. $\nu$ is near the theoretically predicted $-2$ for all combinations of $\ell_o$ and $\chi$.}}}
	\label{sfig:spectraerror}
\end{figure}

As discussed in the main text, we measure Eq.~\eqref{seq:fluctspectrum} from particle-based simulation and plotted the results in Fig. 3.
We fit the measured fluctuation spectra to the power law form $\langle |h(k)|^2 \rangle = K_s k^{\nu}$ in order to extract a scaling with $k$.
We repeated this calculation after splitting all simulation data into five evenly sized time epochs and calculated the standard deviation of the $\nu$ fit between the five epochs.
The average result of $\nu$ and associated errorbars are plotted in Fig.~\ref{sfig:spectraerror}.
\subsection{\label{sec:powerspec}Power Spectrum}
In order to determine dynamical information and the role of $\nu_{\rm odd}$, we now solve for the power spectrum. 
We will do so by Fourier transforming in time:
\begin{equation}
    i\omega \zeta_{\rm eff} h = -|k|^3\upgamma_{\rm cw} h + ik|k|\nu_{\rm odd} h + \tilde{\chi}^{\rm iso} + \tilde{\chi}^{\rm aniso},
\end{equation}
where the noise correlations of $\tilde{\chi}$ are the same as their time-space counterparts but with $\delta(t-t')$ switched with $\delta(\omega+\omega')$.
We can now rearrange the time Fourier transformed evolution equation to find:
\begin{equation}
    h(k,\omega)\left[|k|^3\upgamma_{\rm cw} + i\left(\omega\zeta_{\rm eff} + k|k|\nu_{\rm odd}\right)\right] = \tilde{\chi}^{\rm iso} + \tilde{\chi}^{\rm aniso}.
\end{equation}
Now we multiply both sides of the above equation by its own complex conjugate $(h(-k,-\omega))$ and average over the noise, resulting in:
\begin{multline}
    \langle |h(k,\omega)|^2 \rangle\left[|k|^6\upgamma_{\rm cw}^2 + \left(\omega\zeta_{\rm eff} + k|k|\nu_{\rm odd}\right)^2\right] \\= \langle\tilde{\chi}^{\rm iso}(k,\omega) \tilde{\chi}^{\rm iso}(-k,-\omega) \rangle + \langle\tilde{\chi}^{\rm aniso}(k,\omega) \tilde{\chi}^{\rm aniso}(-k,-\omega) \rangle.\label{seq:powerspecalmost}
\end{multline}
Substituting the noise correlations [Eqs.~\eqref{seq:isonoisecorr} and~\eqref{seq:anisonoisecorr}] into Eq.~\eqref{seq:powerspecalmost} results in:
\begin{equation}
    \langle |h(k,\omega)|^2 \rangle\left[|k|^6\upgamma_{\rm cw}^2 + \left(\omega\zeta_{\rm eff} + k|k|\nu_{\rm odd}\right)^2\right] = 2|k|k_BT^{\rm act}\zeta_{\rm eff}L\Gamma + \zeta_{\rm eff} L\mathcal{A}(k),
\end{equation}
and in the limit $k\ell_o<<1$ where $\mathcal{A}$ is negligible the power spectrum reduces to:
\begin{equation}
    \langle |h(k,\omega)|^2 \rangle = \frac{2|k|k_BT^{\rm act}\zeta_{\rm eff}L\Gamma }{|k|^6\upgamma_{\rm cw}^2 + \left(\omega\zeta_{\rm eff} + k|k|\nu_{\rm odd}\right)^2}
\end{equation}

\subsection{\label{sec:distribution}Distribution}
We now wish to determine whether the linearized interfacial dynamics exhibit surface area minimizing statistics in the steady-state.
We do so by solving for the Fokker-Planck equation corresponding to Eq.~\eqref{seq:interfacelangevin}.
Breaking Eq.~\eqref{seq:interfacelangevin} into real and imaginary parts, recognizing that the noise variance is real, and following the arguments of Zwanzig~\cite{Zwanzig2001NonequilibriumMechanics} results in:
\begin{equation}
    \frac{\partial P[\mathbf{h}]}{\partial t}= \frac{\partial}{\partial \mathbf{h}}\cdot\left[\begin{pmatrix}k^2\upgamma_{\rm cw} & k\nu_{\rm odd} \\ -k\nu_{\rm odd}  & k^2\upgamma_{\rm cw} \end{pmatrix}\cdot\begin{pmatrix}\text{Re}\left(h\right) \\ \text{Im}\left(h\right)\end{pmatrix}P[\mathbf{h}] + \begin{pmatrix}\left(Lk_BT^{\rm act} + \frac{\mathcal{A}}{2|k|}\right)\frac{\partial P}{\partial \text{Re}\left(h\right)} \\ 0\end{pmatrix} \right],
    \label{seq:interfacefokkerplancksub}
\end{equation}
where we are expressing $\mathbf{h} \equiv [\text{Re}\left(h\right),\text{Im}\left(h\right)]$.
We now demand that the \textit{steady-state} solution has no probabilistic flux at the boundaries, so our coupled PDE can be reduced to a system of equations, one of which being:
\begin{equation}
    \left(-k\nu_{\rm odd}\text{Re}\left(h\right) + k^2\upgamma_{\rm cw}\text{Im}\left(h\right)\right)P[\mathbf{h}] = 0.
\end{equation}
This implies that at steady-state,
\begin{equation}
    \text{Im}\left(h\right) = \frac{\nu_{\rm odd}\text{Re}\left(h\right)}{k\upgamma_{\rm cw}}.
\end{equation}
We can substitute in this steady-state relation to the remaining equation, which is now an ordinary differential equation:
\begin{equation}
    \left(k^2\upgamma_{\rm cw} + \frac{\nu_{\rm odd}^2}{\upgamma_{\rm cw}}\right)\text{Re}\left(h\right) P[\mathbf{h}] + \left(Lk_BT^{\rm act} + \frac{\mathcal{A}}{2|k|}\right) \frac{\partial P[\mathbf{h}]}{\partial \text{Re}\left(h\right)} = 0
\end{equation}
The solution of this differential equation is:
\begin{equation}
    P[\mathbf{h}] \sim \exp{\left[-\frac{\frac{1}{2}\left(k^2\upgamma_{\rm cw} + \frac{\nu_{\rm odd}^2}{\upgamma_{\rm cw}}\right)\text{Re}\left(h\right)^2}{Lk_BT^{\rm act} + \frac{\mathcal{A}}{2|k|}}\right]},
\end{equation}
and substituting in the steady-state relationship between the imaginary and real components of $h$ allows one to find:
\begin{equation}
    P[\mathbf{h}] \sim \exp{\left[-\frac{k^2\upgamma_{\rm cw} |\mathbf{h}|^2}{2Lk_BT^{\rm act} + \frac{\mathcal{A} }{|k|}}\right]}.
\end{equation}
In the limit $k\ell_o << 1$, where $\mathcal{A}$ was found to be negligible~\cite{Langford2023}, the distribution takes on the simple Boltzmann form:
\begin{equation}
    P[\mathbf{h}] \sim \exp{\left[-\frac{k^2\upgamma_{\rm cw} |\mathbf{h}|^2}{2Lk_BT^{\rm act} }\right]},
\end{equation}
recovering a surface-area minimizing Boltzmann distribution even in the presence of chirality.
 \newpage
\section{\label{sec:media}Supporting Media}
Each video listed below displays the instantaneous interface determined from a Brownian dynamics simulation of cABP phase separation at a specified activity and chirality. 
The semi-transparent magenta region corresponds to the liquid phase while the uncolored region corresponds to the gas phase.
In each video, the instantaneous interface was computed using a cubic grid of points with a uniform spacing of $0.89~d_{\rm_{hs}}$ and a coarse-graining length of $\xi/d_{\rm{hs}} = 1.78$. 
In each video, the instantaneous interface was computed using the algorithm developed by Patch \textit{et al.}~\cite{Patch2018}.
This method first employs a clustering algorithm to determine the particles within the liquid phase, then sorts each liquid particle into bins by its $y$ coordinate. 
The $x$ coordinate of the three particles with the highest $x$ coordinate in each bin were then averaged, giving the location of the right interface corresponding to the $y$ value associated with the bin.
The $x$ coordinate of the three particles with the lowest $x$ coordinate in each bin were then averaged, giving the location of the left interface corresponding to the $y$ value associated with the bin.
Each frame is separated by a duration of $4.45~d_{\rm hs}/U_o$ and the videos are played at a rate of $100$ frames per second.
All videos are publicly available at the following URL:
\newline
\url{https://berkeley.box.com/s/7befhar83g2nnl3ci8sdt7rkc6hb3aqh}
\newline
\begin{itemize}
    \item
    \text{[}interface\_75LR\_0chi.mp4\text{]}:\newline 
    Instantaneous interface dynamics with an activity of $\ell_o/d = 75$ and a chirality of $\chi = 0$. 
    \item
    \text{[}interface\_75LR\_025chi.mp4\text{]}:\newline 
    Instantaneous interface dynamics with an activity of $\ell_o/d = 75$ and a chirality of $\chi = 0.25$.
    \item
    \text{[}interface\_75LR\_05chi.mp4\text{]}:\newline 
    Instantaneous interface dynamics with an activity of $\ell_o/d = 75$ and a chirality of $\chi = 0.5$.
    \item
    \text{[}interface\_75LR\_075chi.mp4\text{]}:\newline 
    Instantaneous interface dynamics with an activity of $\ell_o/d = 75$ and a chirality of $\chi = 0.75$. 
    \item
    \text{[}interface\_100LR\_0chi.mp4\text{]}:\newline 
    Instantaneous interface dynamics with an activity of $\ell_o/d = 100$ and a chirality of $\chi = 0$. 
    \item
    \text{[}interface\_100LR\_025chi.mp4\text{]}:\newline 
    Instantaneous interface dynamics with an activity of $\ell_o/d = 100$ and a chirality of $\chi = 0.25$.
    \item
    \text{[}interface\_100LR\_05chi.mp4\text{]}:\newline 
    Instantaneous interface dynamics with an activity of $\ell_o/d = 100$ and a chirality of $\chi = 0.5$.
    \item
    \text{[}interface\_100LR\_075chi.mp4\text{]}:\newline 
    Instantaneous interface dynamics with an activity of $\ell_o/d = 100$ and a chirality of $\chi = 0.75$.
    \item
    \text{[}interface\_125LR\_0chi.mp4\text{]}:\newline 
    Instantaneous interface dynamics with an activity of $\ell_o/d = 125$ and a chirality of $\chi = 0$. 
    \item
    \text{[}interface\_125LR\_025chi.mp4\text{]}:\newline 
    Instantaneous interface dynamics with an activity of $\ell_o/d = 125$ and a chirality of $\chi = 0.25$.
    \item
    \text{[}interface\_125LR\_05chi.mp4\text{]}:\newline 
    Instantaneous interface dynamics with an activity of $\ell_o/d = 125$ and a chirality of $\chi = 0.5$.
    \item
    \text{[}interface\_125LR\_075chi.mp4\text{]}:\newline 
    Instantaneous interface dynamics with an activity of $\ell_o/d = 125$ and a chirality of $\chi = 0.75$. 
\end{itemize}